\documentclass[11pt]{article}
\usepackage{amsmath, amsfonts, amsthm, amssymb}
\usepackage{hyperref}

\newcommand{\abs}[1]{{\lvert #1 \rvert}}
\newcommand{\norm}[1]{{\lVert #1 \rVert}}
\newcommand{\set}[1]{{\lbrace{#1}\rbrace}}

\newcommand{\tensor}{\otimes}
\newcommand{\Tensor}{\bigotimes}

\newcommand{\RR}{\mathbb{R}}
\newcommand{\CC}{\mathbb{C}}
\newcommand{\eps}{\varepsilon}
\newcommand{\calH}{\mathcal{H}}
\newcommand{\Atil}{\widetilde{A}}

\newcommand{\Ytil}{\widetilde{Y}}
\newcommand{\vecp}{\vec{p}}
\newcommand{\vecq}{\vec{q}}
\newcommand{\vecu}{\vec{u}}
\newcommand{\vecd}{\vec{d}}
\newcommand{\calD}{\mathcal{D}}
\newcommand{\calE}{\mathcal{E}}
\newcommand{\calF}{\mathcal{F}}
\newcommand{\Mtil}{\widetilde{M}}
\newcommand{\Wtil}{\widetilde{W}}
\newcommand{\Ctil}{\widetilde{C}}
\newcommand{\Util}{\widetilde{U}}
\newcommand{\Vtil}{\widetilde{V}}
\newcommand{\Ztil}{\widetilde{Z}}
\newcommand{\Lambdatil}{\widetilde{\Lambda}}
\newcommand{\Ltil}{\widetilde{L}}
\newcommand{\Ktil}{\widetilde{K}}
\newcommand{\calEtil}{\widetilde{\calE}}

\DeclareMathOperator{\EE}{\mathbb{E}}
\DeclareMathOperator{\Tr}{tr}

\theoremstyle{plain}
\newtheorem{definition}{Definition}[section]
\newtheorem{theorem}[definition]{Theorem}
\newtheorem{proposition}[definition]{Proposition}
\newtheorem{lemma}[definition]{Lemma}
\newtheorem{corollary}[definition]{Corollary}

\begin{document}

\title{Privacy Amplification in the Isolated Qubits Model}
\author{Yi-Kai Liu\\
Applied and Computational Mathematics Division\\
National Institute of Standards and Technology\\
Gaithersburg, MD, USA\\
yi-kai.liu@nist.gov}
\date{\today}
\maketitle

\abstract{
Isolated qubits are a special class of quantum devices, which can be used to implement tamper-resistant cryptographic hardware such as one-time memories (OTM's).  Unfortunately, these OTM constructions leak some information, and standard methods for privacy amplification cannot be applied here, because the adversary has advance knowledge of the hash function that the honest parties will use.  

In this paper we show a stronger form of privacy amplification that solves this problem, using a \textit{fixed} hash function that is secure against all possible adversaries in the isolated qubits model.  This allows us to construct single-bit OTM's which only leak an exponentially small amount of information.  

We then study a natural generalization of the isolated qubits model, where the adversary is allowed to perform a polynomially-bounded number of entangling gates, in addition to unbounded local operations and classical communication (LOCC).  We show that our technique for privacy amplification is also secure in this setting.

}


\section{Introduction}

Can one build tamper-resistant cryptographic hardware whose security is based on the laws of quantum mechanics?  This is a natural question, as there are many unusual phenomena in quantum mechanics, such as the impossibility of cloning an unknown quantum state, which seem relevant to cryptography.  However, despite these encouraging signs, it turns out that many common cryptographic functionalities, such as bit commitment and oblivious transfer (with information-theoretic security), \textit{cannot} be implemented in a quantum world \cite{nogo1, nogo2, nogo3, nogo4}.  

Recently, there has been progress using a different approach to this problem, called the ``isolated qubits model'' \cite{liu-crypto2014, liu-itcs2014}.  Isolated qubits are qubits with long coherence times, which can only be accessed using single-qubit gates and measurements; entangling operations are forbidden.  Thus, in the isolated qubits model, one assumes an additional restriction on what the adversary can do.  Formally, the adversary is only allowed to perform local operations and classical communication, or LOCC, where ``local operations'' are operations on single qubits, and ``classical communication'' refers to communication between the qubits.  (Likewise, honest parties are also restricted to LOCC.  Furthermore, while the adversary can perform an unbounded number of operations, all honest parties must run in polynomial time.)  Isolated qubits can be viewed as special-purpose quantum devices, which can implement functionalities such as oblivious transfer that are not possible using quantum mechanics alone.  Isolated qubits could conceivably be implemented using solid-state nuclear spins, such as quantum dots or nitrogen vacancy centers \cite{saeedi-2013, dreau-2013}.  

Using isolated qubits, there are natural candidate constructions that lead to a variety of tamper-resistant cryptographic hardware.  The first step is to construct \textit{one-time memories} (OTM's) \cite{liu-crypto2014}.  Intuitively, a one-time memory is a device that does non-interactive oblivious transfer, i.e., Alice programs the device with two messages $s$ and $t$, then gives the device to Bob, who can choose to read either $s$ or $t$ (but not both).  

Using one-time memories, one can then construct \textit{one-time programs} \cite{GKR, goyal, bellare, broadbent}, which are useful for program obfuscation, access control and copy protection.  A one-time program is a program that can be run only once, and hides its internal state.  More precisely, Alice chooses some circuit $C$, compiles it into a one-time program, and gives it to Bob; Bob then chooses an input $x$, runs the one-time program, and learns the output of the computation $C(x)$; but Bob learns nothing else, and cannot run the program on another input.  

Unfortunately it is not yet possible to prove the security of these one-time programs in the isolated qubits model.  This is because the proof of security for the one-time memories in \cite{liu-crypto2014} is not strong enough --- it allows some extra information to \textit{leak} to the adversary, which can cause problems when the one-time memories are used as part of a larger construction.  

In this paper we address the issue of information leakage, by developing a privacy amplification technique that works in the isolated qubits model.  By combining this privacy amplification technique with the leaky one-time memories from \cite{liu-crypto2014}, we obtain new one-time memories that only leak an exponentially small amount of information.  These new one-time memories store single bits rather than strings, but these can also be plugged into known constructions for one-time programs \cite{goyal}.  This removes one of the main obstacles to constructing provably-secure one-time programs.

\subsection{Privacy amplification}

The candidate construction for one-time memories in \cite{liu-crypto2014} was proven to satisfy a ``leaky'' definition of security, where up to a constant fraction of the bits of the messages could be leaked to the adversary.  This notion of security was not as strong as one would have liked, but on the positive side, the adversary's uncertainty was expressed in terms of the smoothed min-entropy, which suggested that the leakage problem might be addressed using some kind of privacy amplification.

However, there is an obstacle to using privacy amplification with our one-time memories.  Usually, in privacy amplification, the adversary has partial information about some string $s$ (while the honest parties have complete knowledge of $s$).  Then the honest parties choose a random seed $q$, and apply a hash function $F_q$ to produce a shorter string $F_q(s)$, which will be almost completely unknown to the adversary.  This works provided that the random seed $q$ is chosen independent of the adversary's actions.  

But in the case of our one-time memories, all the information needed to decode the messages --- including the random seed $q$ --- must be provided at the beginning, before the adversary decides how to attack the OTM (i.e., what measurement to perform on the qubits).  Thus the adversary's attack can depend on $q$, and so standard methods of privacy amplification may not be secure.

We show a variant of privacy amplification which uses a \textit{fixed} hash function $F$ (without a random seed), and is secure in the isolated qubits model.  Intuitively, this relies on two ideas.  First, we use a stronger family of hash functions, namely $r$-wise independent functions, where $r$ grows polynomially in the security parameter $k$.  These $r$-wise independent functions can be computed efficiently, but they behave more like truly random functions, in that they satisfy large-deviation bounds, similar to Hoeffding's inequality \cite{vadhan, bellare-rompel, srinivasan}.  

Second, we exploit the fact that the only way for the adversary to learn about $s$ is by performing LOCC measurements on the qubits that encode $s$.  Rather than considering all possible LOCC measurement strategies, which are represented by decision trees, we consider all possible LOCC measurement outcomes, which are represented by POVM elements.\footnote{POVM elements are defined in Section \ref{sec-meas}, but we do not require these formal definitions here.}  Due to the LOCC restriction, these POVM elements are tensor products of single-qubit operators.  So there are not too many of them.  Say we discretize the set of possible measurement outcomes, with some fixed resolution.\footnote{Formally, we consider an $\eps$-net, as defined in Section \ref{sec-notation}.}  Then the number of LOCC measurement outcomes grows exponentially with the number of qubits; this is in contrast with the number of \textit{entangled} measurement outcomes, which grows doubly-exponentially with the number of qubits.  Hence we can use a union bound over the set of all LOCC measurement outcomes.

We do privacy amplification as follows.  We first choose a hash function $F$ from an $r$-wise independent family.  We then fix $F$ permanently, and announce it to the adversary.  We claim that, with high probability over the choice of $F$, this privacy amplification scheme will be secure against \textit{all} possible LOCC adversaries, i.e., every adversary who uses LOCC measurements and gains at most partial information about the string $s$, will still have very little information about $F(s)$.  

The proof uses a covering argument over the set of all LOCC measurement outcomes.  First, fix some particular LOCC measurement outcome $M$.  Let $S$ be the random variable representing the string $s$, and suppose the hash function $F$ outputs a single bit $F(S)$.  One can calculate the bias of the bit $F(S)$, conditioned on having observed outcome $M$, as follows:
\begin{equation}
\EE_S((-1)^{F(S)} \,|\, M) = \sum_s (-1)^{F(s)} \Pr(S=s \,|\, M).
\end{equation}
(Here $\EE_S$ denotes the expectation value obtained by averaging over $S$.)  

We want to show that $\EE_S((-1)^{F(S)} \,|\, M)$ is small.  Notice that $\EE_S((-1)^{F(S)} \,|\, M)$ is a linear combination of terms $(-1)^{F(s)}$, where each $F(s)$ is a random variable describing the initial choice of the hash function $F$.  We can use Hoeffding-like inequalities to show that, with high probability over the choice of $F$, $\EE_S((-1)^{F(S)} \,|\, M)$ is sharply concentrated around 0.  This will work provided that $\sum_s \Pr(S=s \,|\, M)^2$ is small, which follows since the Renyi entropy $H_2(S|M)$ (or the smoothed min-entropy $H_\infty^\eps(S|M)$) are large, which holds since the adversary has at most partial information about $S$.  Thus, one can conclude that, for a fixed LOCC measurement outcome $M$, with high probability over the choice of $F$, $\EE_S((-1)^{F(S)} \,|\, M)$ is small, i.e., privacy amplification succeeds.  

Finally, one uses the union bound over all LOCC measurement outcomes $M$.  This shows that, with high probability over the choice of $F$, for all LOCC measurement outcomes $M$, privacy amplification succeeds.  This completes the proof.

The above sketch shows privacy amplification for a single string $s$, but a similar technique can be applied to an OTM that stores two strings $s$ and $t$.  Here one applies two hash functions $F$ and $G$, which output a pair of bits $F(s)$ and $G(t)$.  Now there is an additional complication, since the adversary has the possibility of learning information about the \textit{correlations} between $F(s)$ and $G(t)$.  To address this issue, one needs to bound the quantity 
\begin{equation}
\EE_{ST}((-1)^{F(S)+G(T)} \;|\; M) = \sum_{st} (-1)^{F(s)} (-1)^{G(t)} \Pr(S=s, T=t \;|\; M).
\end{equation}
This is a quadratic function of the random variables $(-1)^{F(s)}$ and $(-1)^{G(t)}$, describing the initial choices of the hash functions $F$ and $G$.  This can be bounded using the Hanson-Wright inequality \cite{hanson-wright, hanson-wright-RV}, adapted for $r$-wise independent variables using the techniques of \cite{bellare-rompel, srinivasan}.

Formally, this shows a reduction from an almost-perfect single-bit OTM to a leaky string-OTM.  (That is, given an OTM that stores two strings and leaks a constant fraction of the information, one can construct an OTM that stores two bits and leaks an exponentially small amount of information.)  By combining with the results of \cite{liu-crypto2014}, we get almost-perfect single-bit OTM's in the isolated qubits model.

\subsection{Beyond the isolated qubits model}

Next, we study a generalization of the isolated qubits model, where the adversary is allowed to perform a polynomially-bounded number of 2-qubit entangling gates, in addition to unbounded LOCC operations.  More precisely, this model is specified by a constant $c \geq 0$, and a ``depth'' parameter $d$, which can grow with the security parameter $k$, as long as $d \leq k^c$; then this model allows the adversary to apply quantum circuits of depth $d$ containing 2-qubit gates combined with unbounded LOCC operations.  (Honest parties are still restricted to polynomial-time LOCC.)  This model may be a more accurate description of real solid-state qubits, where one can perform noisy entangling gates, but the accumulation of noise makes it difficult to entangle large numbers of qubits at once.

It is an interesting open problem to construct OTM's that are secure in this model.  We show that our reduction from almost-perfect single-bit OTM's to leaky string-OTM's still works in this setting.  More precisely, for any constant $c \geq 0$, and any depth $d \leq k^c$, we show a variant of our reduction, whose efficiency is polynomial in $d$, that remains secure in this depth-$d$ model.  The proof uses the same ideas as before.  

Unfortunately, the leaky string-OTM's from \cite{liu-crypto2014} are not known to be secure in this setting.  Nonetheless we believe it should be possible to construct leaky string-OTM's in this depth-$d$ model, for at least some super-constant values of $d$, for the following intuitive reason:  in order to break the leaky string-OTM's from \cite{liu-crypto2014}, one has to break a particular version of Wiesner's conjugate coding scheme \cite{wiesner}, and this requires running a classical decoding algorithm on a quantum superposition of inputs, which requires applying a quantum circuit with a certain minimum number of entangling gates.

\subsection{Discussion}

\textbf{Related work:}
This paper builds on recent work on non-interactive one-time memories in the isolated qubits model \cite{liu-crypto2014, liu-itcs2014}.  Some similar ideas have been investigated in connection with other cryptographic tasks, such as bit commitment, quantum money and password-based identification \cite{salvail, pastawski, all-but-one}.  There is also a related line of work on LOCC state discrimination, involving ``nonlocality without entanglement'' and data-hiding states \cite{nlwe99, nlwe12, qdh02, eggeling02}.  

Deterministic privacy amplification has also been studied for other cryptographic tasks, such as secret key distribution based on causality constraints \cite{masanes}.  Our results can also be compared to earlier work on deterministic extractors for special classes of random sources, as well as exposure-resilient cryptography and leakage-resilient cryptography \cite{trevisan-vadhan-2000, kamp-zuckerman-2006, gabizon-2011, AGV, naor-segev}.  However, these earlier works considered classical adversaries, with various kinds of restrictions; our result, with a quantum adversary restricted to (unbounded) LOCC operations, seems to be new.

\textbf{Open problems:}
The overall goal of this work is to construct one-time programs whose security is based on the properties of realistic physical devices.  One-time memories and the isolated qubits model are useful steps along the way to achieving this goal, but there remain several open problems.  

First, can one prove that these one-time memories satisfy a sufficiently strong notion of security, so that they can be composed to build one-time programs?  Privacy amplification is helpful, but there may be other issues that affect the security of more complicated protocols, such as the adversary's ability to wait until later stages of the protocol before performing any measurements.  

Second, can one modify the isolated qubits model so that it matches more closely the properties of real solid-state qubits, e.g., by allowing a limited number of entangling operations?  Our model involving bounded-depth quantum circuits is one step in this direction.


\section{Preliminaries}

\subsection{Notation, $\eps$-nets}
\label{sec-notation}

For any two matrices $A$ and $B$, we write $A \preceq B$ if and only if $B-A$ is positive semidefinite.  We let $\norm{A}$ denote the operator norm, $\norm{A}_{\Tr}$ denote the trace norm, and $\norm{A}_F$ denote the Frobenius norm.  For any vector $v$, we let $\norm{v}_p$ denote the $\ell_p$ norm of $v$.  For any two probability densities $P$ and $Q$, we let $\norm{P-Q}_1$ denote the $\ell_1$ distance between them.

We write $\Pr[\calE]$ to denote the probability of an event $\calE$.  We write $\EE[X]$ to denote the expectation value of a random variable $X$.  In some cases we write $\Pr_X[\cdot]$ or $\EE_X[\cdot]$ to emphasize that we are considering probabilities associated with a random variable $X$.  We write $P_{X|Y}$ to denote a probability density function $P_{X|Y}(x|y) = \Pr[X=x \,|\, Y=y]$.  In some cases we abuse this notation, e.g., if $\calE$ is an event, we write $P_{\calE X|Y}(x|y) = \Pr[\calE,\, X=x \,|\, Y=y]$.  Also, if $\calE$ is an event, we let $1_\calE$ be the indicator random variable for $\calE$, which equals 1 when the event $\calE$ happens, and equals 0 otherwise.

Suppose $E$ is a subset of some normed space, with norm $\norm{\cdot}$.  Let $\eps > 0$.  We say that $\tilde{E}$ is an \textit{$\eps$-net} for $E$ if $\tilde{E} \subset E$, and for every $x \in E$, there exists some $y \in \tilde{E}$ such that $\norm{x-y} \leq \eps$.

\subsection{Quantum measurements}
\label{sec-meas}

A quantum state is described by a \textit{density matrix} $\rho \in \CC^{d\times d}$ with $\rho \succeq 0$ and $\Tr(\rho) = 1$.  A quantum measurement can be described by a \textit{completely-positive trace-preserving map} $\calE:\: \CC^{d\times d} \rightarrow \CC^{d\times d}$, which can be written in the form $\calE(\rho) = \sum_i K_i \rho K_i^\dagger$, where the $K_i \in \CC^{d\times d}$ are called \textit{Kraus operators} and $\sum_i K_i^\dagger K_i = I$.  Given a state $\rho$, the measurement returns outcome $i$ with probability $\Tr(K_i \rho K_i^\dagger)$, in which case the post-measurement state is given by $K_i \rho K_i^\dagger / \Tr(K_i \rho K_i^\dagger)$.

A measurement outcome can also be described by a \textit{POVM element},\footnote{POVM refers to positive operator-valued measure, though we will not need to use this concept here.} that is, a matrix $M \in \CC^{d\times d}$ with $0 \preceq M \preceq I$.  Given a state $\rho$, the probability that a measurement returns outcome $M$ is given by $\Tr(M\rho)$.  (In the example in the previous paragraph, the outcome $i$ is described by the POVM element $K_i^\dagger K_i$.)


\subsection{LOCC and separable measurements}
\label{sec-locc}

In the isolated qubits model, qubits are only accessible via \textit{local operations and classical communication} (LOCC), that is, one can perform single-qubit quantum operations, and use classical information (obtained by measuring one qubit) to choose what operation to perform on another qubit.  LOCC strategies can thus be represented by decision trees, where each vertex corresponds to a single-qubit operation, and each edge corresponds to a possible (classical) outcome of that operation \cite{liu-crypto2014, liu-itcs2014}.  

A measurement on $m$ qubits is called \textit{separable} if it can be written in the form $\calE:\: \rho \mapsto \sum_i K_i \rho K_i^\dagger$, where each operator $K_i$ is a tensor product of $m$ single-qubit operators, $K_i = K_{i,1} \tensor K_{i,2} \tensor \cdots \tensor K_{i,m}$.  It is easy to see that any LOCC measurement is separable \cite{horodecki}.  

\subsection{Smoothed min-entropy}

We recall the definition of the smoothed conditional min-entropy:  
\begin{equation}
H_\infty^\eps(X|Y) = \max_{\calE:\; \Pr(\calE) \geq 1-\eps} \min_{x,y} \Bigl[ -\lg \bigl[ P_{\calE X|Y}(x|y) \bigr] \Bigr], 
\end{equation}
where the maximization is over all events $\calE$ (defined by the conditional probabilities $P_{\calE|XY}$) such that $\Pr(\calE) \geq 1-\eps$.  Note that a lower-bound of the form $H_\infty^\eps(X|Y) \geq h$ implies that there exists an event $\calE$ with $\Pr(\calE) \geq 1-\eps$ such that, for all $x$ and $y$, $\Pr[\calE,X=x|Y=y] \leq 2^{-h}$.  

We will need the following ``entropy splitting lemma,'' which appeared in \cite{damgard}.  Intuitively, this says that if $X_0$ and $X_1$ together have min-entropy at least $\alpha$, then at least one of them (indicated by the random variable $C$) must have min-entropy at least $\alpha/2$.
\begin{proposition}
\label{prop-rooster}
Let $\eps \geq 0$, and let $X_0$, $X_1$ and $Z$ be random variables (which may be over different alphabets) such that $H_\infty^\eps (X_0,X_1 \,|\, Z) \geq \alpha$.  Then there exists a random variable $C$ taking values in $\set{0,1}$ such that 
\begin{equation}
H_\infty^{\eps+\eps'} (X_{1-C} \,|\, Z,C) \geq \tfrac{1}{2}\alpha - 1 - \lg(\tfrac{1}{\eps'}) \qquad (\text{for any } \eps' > 0).
\end{equation}
\end{proposition}

\subsection{Leaky string-OTM's}

The main result from \cite{liu-crypto2014} was a construction of a leaky string-OTM (which stores two strings, and leaks at most a constant fraction of the information) in the isolated qubits model.  Here we state this result using slightly different language --- in particular, we state the result in terms of ``$\delta$-non-negligible'' measurement outcomes, whereas in \cite{liu-crypto2014} this terminology was used in the proof but not in the statement of the theorem.

\begin{definition}
\label{def-delta-non-negl}
For any quantum state $\rho \in \CC^{d\times d}$, and any $\delta > 0$, we say that a measurement outcome (POVM element) $M \in \CC^{d\times d}$ is \textit{$\delta$-non-negligible} if $\Tr(M\rho) \geq \delta \Tr(M) / d$.  
\end{definition}

Intuitively, these are the only measurement outcomes we need to consider in our security proof, as the total probability contributed by all the other ``$\delta$-negligible'' measurement outcomes is never more than $\delta$.  To see this, consider any measurement, which can be described by a collection of POVM elements $\set{M_z \;|\; z = 1,2,\ldots}$ such that $\sum_z M_z = I$.  Say we perform this measurement on some state $\rho$, and let $Z$ be the random variable representing the outcome of the measurement (so $Z$ takes values $M_1,M_2,\ldots$).  Then the total probability of observing a $\delta$-negligible measurement outcome is at most $\delta$:
\begin{equation}
\label{eqn-delta-negl}
\Pr[\text{outcome $Z$ is $\delta$-negligible}]
 = \sum_{\text{$z$ : $M_z$ is $\delta$-negl.}} \Tr(M_z\rho)
 < \sum_{\text{$z$ : $M_z$ is $\delta$-negl.}} \delta \Tr(M_z) / d
 \leq \delta.
\end{equation}

We now restate the main result from \cite{liu-crypto2014}:
\begin{theorem}\label{thm-night-hawk}
For any $k \geq 2$, and for any small constant $0<\mu \ll 1$, there exists an OTM construction that stores two messages $s,t \in \set{0,1}^\ell$, where $\ell = \Theta(k^2)$, and has the following properties:
\begin{enumerate}
\item Correctness and efficiency:  there are honest strategies for programming the OTM with messages $s$ and $t$, and for reading either $s$ or $t$, using only LOCC operations, and time polynomial in $k$.
\item ``Leaky'' security:  Let $\delta_0 > 0$ be any constant, and set $\delta = 2^{-\delta_0 k}$.  Suppose the messages $s$ and $t$ are chosen independently and uniformly at random in $\set{0,1}^\ell$.  For any LOCC adversary, and any separable\footnote{Note that this includes LOCC measurement outcomes as a special case.} measurement outcome $M$ that is $\delta$-non-negligible, we have the following security bound:
\begin{equation}
\begin{split}
H_\infty^{\eps}(S,T|Z=M)
&\geq (\tfrac{1}{2}-\mu) \, \ell - \delta_0 k.
\end{split}
\end{equation}
Here $S$ and $T$ are the random variables describing the two messages, $Z$ is the random variable representing the adversary's measurement outcome, and we have $\eps \leq \exp(-\Omega(k))$.
\end{enumerate}
\end{theorem}

\subsection{Ideal bit-OTM's}

We now define security for an ``ideal'' OTM that stores two bits $a_0, a_1 \in \set{0,1}$.  Note that there is a subtle point with defining security:  while the OTM should hide at least one of the messages $(a_0, a_1)$, which one remains hidden may depend on the adversary's actions in a complicated way.  Our definition of security asserts that, conditioned on the adversary's measurement outcome, there exists a binary random variable $C$ that indicates which of the two messages remains hidden.  (For example, $C$ appears naturally when one uses the entropy splitting lemma, Prop. \ref{prop-rooster}.)  

Formally, we let $A_0$ and $A_1$ be random variables representing the two messages.  Our security definition asserts that the message $A_C$ is nearly uniformly distributed, even given knowledge of the other message $A_{1-C}$, the value of $C$, and the adversary's measurement outcome.

\begin{definition}
\label{def-security}
We say that a single-bit OTM construction is secure if the following holds:
Let $k\geq 1$ be a security parameter.  Suppose the OTM is programmed with two messages $a_0, a_1 \in \set{0,1}$ chosen uniformly at random.  Consider any LOCC adversary, and let $Z$ be the random variable representing the results of the adversary's measurements.  Then there exists a random variable $C$, which takes values in $\set{0,1}$, such that:
\begin{equation}
\begin{split}
\norm{P_{A_C A_{1-C} C Z} - U_{A_C} \times P_{A_{1-C} C Z}}_1 \leq 2^{-\Omega(k)},
\end{split}
\end{equation}
where $P_{A_C A_{1-C} C Z}$ denotes the probability density on the random variables $(A_C, A_{1-C}, C, Z)$, $P_{A_{1-C} C Z}$ denotes the marginal probability density on $(A_{1-C}, C, Z)$, and $U$ denotes the uniform distribution on $\set{0,1}$.
\end{definition}

We remark that this security guarantee involves the adversary's measurement outcome $Z$, which is \textit{classical} rather than quantum information.  While this may seem like an artificial restriction on the adversary, we argue that it is simply a natural consequence of the isolated qubits model.  By definition, the adversary is unable to perform any entangling operations on the isolated qubits contained in the OTM; thus the only way the adversary can access those qubits is by performing a measurement, and converting the quantum state into a classical measurement outcome.

\subsection{$t$-wise independent hash functions}

Let $\calH$ be a collection of functions $h$ that map $\set{1,\ldots,N}$ to $\set{1,\ldots,M}$.  Let $t \geq 1$ be an integer.  Let $H$ be a function chosen uniformly at random from $\calH$; then this defines a collection of random variables $\set{H(x) \;|\; x = 1,\ldots,N}$.  We say that $\calH$ is \textit{$t$-wise independent} if for all subsets $S \subset \set{1,\ldots,N}$ of size $\abs{S} \leq t$, the random variables $\set{H(x) \,|\, x \in S}$ are independent and uniformly distributed in $\set{1,\ldots,M}$.

We will use the fact that there exist efficient constructions for $t$-wise independent hash functions, which run in time polynomial in $t$, $\log N$ and $\log M$; see \cite{vadhan} for details.  
\begin{proposition}
For all integers $n \geq 1$, $m \geq 1$ and $t \geq 1$, there exist families of $t$-wise independent functions $\calH = \set{h:\: \set{0,1}^n \rightarrow \set{0,1}^m}$, such that sampling a random function in $\calH$ takes $t \cdot \max\set{n,m}$ random bits, and evaluating a function in $\calH$ takes time $\text{poly}(n,m,t)$.
\end{proposition}

We will use the following large-deviation bound for sums of $t$-wise independent random variables.  This is a slight variant of results in \cite{bellare-rompel} (see also \cite{srinivasan}); we sketch the proof in Appendix \ref{app-starfish}.
\begin{proposition}
\label{prop-kite}
Let $t \geq 2$ be an even integer, and let $\calH$ be a family of $t$-wise independent functions that map $\set{1,\ldots,N}$ to $\set{0,1}$.  Fix some constants $a_1,\ldots,a_N \in \RR$.  Let $H$ be a function chosen uniformly at random from $\calH$, and define the random variable 
\begin{equation}
Y = \sum_{x=1}^N (-1)^{H(x)} a_x.  
\end{equation}

Then $\EE Y = 0$, and we have the following large-deviation bound:  for any $\lambda > 0$, 
\begin{equation}
\Pr(\abs{Y} \geq \lambda) \leq 2e^{1/(6t)} \sqrt{\pi t} \biggl( \frac{vt}{e\lambda^2} \biggr)^{t/2},
\end{equation}
where $v = \sum_{x=1}^N a_x^2$.
\end{proposition}

We will also use a large-deviation bound for quadratic functions of $2t$-wise independent random variables.  This is based on the Hanson-Wright inequality \cite{hanson-wright} (see also \cite{hanson-wright-RV} for a more modern, slightly stronger result), partially derandomized using the techniques of \cite{bellare-rompel} (see also \cite{srinivasan}).  We sketch the proof in Appendix \ref{app-crayfish}.
\begin{proposition}
\label{prop-crayfish}
Let $t \geq 2$ be an even integer, and let $\calH$ be a family of $2t$-wise independent functions that map $\set{1,\ldots,N}$ to $\set{0,1}$.  Let $A \in \RR^{N\times N}$ be a symmetric matrix, $A^T = A$.  Let $H$ be a function chosen uniformly at random from $\calH$, and define the random variable 
\begin{equation}
S = \sum_{x,y=1}^N A_{xy} \bigl( (-1)^{H(x)} (-1)^{H(y)} - \delta_{xy} \bigr), 
\end{equation}
where $\delta_{xy}$ equals 1 if $x=y$, and equals 0 otherwise.  

Then $\EE S = 0$, and we have the following large-deviation bound:  for any $\lambda > 0$, 
\begin{equation}
\Pr(\abs{S} \geq \lambda) 
\leq 4e^{1/(6t)} \sqrt{\pi t} \biggl( \frac{4 \norm{\Atil}_F^2 t}{e\lambda^2} \biggr)^{t/2} 
   + 4e^{1/(12t)} \sqrt{2\pi t} \biggl( \frac{8 \norm{\Atil} t}{e\lambda} \biggr)^t,
\end{equation}
where $\Atil \in \RR^{N\times N}$ is the entry-wise absolute value of $A$, that is, $\Atil_{xy} = \abs{A_{xy}}$.
\end{proposition}


\section{Privacy amplification for one-time memories using isolated qubits}

Our main result is a reduction from ``ideal'' one-time memories to ``leaky'' one-time memories, in the isolated qubits model.  More precisely, we assume the existence of a ``leaky'' one-time memory $\calD$ that stores two strings $s,t \in \set{0,1}^\ell$, and leaks any constant fraction of the bits of $(s,t)$.  (Such leaky OTM's were constructed previously in \cite{liu-crypto2014}.)  We then construct an ``ideal'' one-time memory $\calD'$ that stores two bits $a_0, a_1 \in \set{0,1}$, and leaks an exponentially small amount of information about either $a_0$ or $a_1$ (so that at least one of the bits $(a_0, a_1)$ remains almost completely hidden).  

Our construction makes use of two functions $F,G:\: \set{0,1}^\ell \rightarrow \set{0,1}$, which are chosen from an $r$-wise independent random ensemble.  (We will specify the value of $r$ later, in the statement of Theorem \ref{thm-main}.)  Once the functions $F$ and $G$ have been chosen, they are fixed permanently, and they become part of the public description of the one-time memory $\calD'$.  (In particular, the adversary may attack $\calD'$ using LOCC strategies that depend on $F$ and $G$.  We show that with high probability over the choice of $F$ and $G$, $\calD'$ is secure against all such attacks.)

We define the ``ideal'' one-time memory $\calD'$ to have the following behavior.  First, one can program $\calD'$ with two messages $a_0, a_1 \in \set{0,1}$.  $\calD'$ implements this functionality in the following way:
\begin{enumerate}
\item Choose $s \in F^{-1}(a_0)$ and $t \in G^{-1}(a_1)$ uniformly at random, e.g., using rejection sampling.\footnote{Choose $s,t \in \set{0,1}^\ell$ uniformly at random, and repeat until one gets $s$ and $t$ that satisfy $F(s) = a_0$ and $G(t) = a_1$.}
\item Program a ``leaky'' one-time memory $\calD$ with the messages $s$ and $t$, and return $\calD$.
\end{enumerate}
Given the device $\calD'$, an honest user can retrieve either $a_0$ or $a_1$ as follows:  
\begin{enumerate}
\item Read either $s$ or $t$ from the device $\calD$, as appropriate.
\item Compute either $a_0 = F(s)$ or $a_1 = G(t)$, as appropriate.
\end{enumerate}

We now prove the correctness and security of these ``ideal'' one-time memories $\calD'$.  
\begin{theorem}
\label{thm-main}
Fix some constants $k_0 \geq 1$, $\theta \geq 1$, $\delta_0 > 0$, $\alpha > 0$ and $\eps_0 > 0$.  

\vskip 10pt

Suppose we have a family of devices $\calD = \set{\calD_k \;|\; k \geq k_0}$, indexed by a security parameter $k \geq k_0$.  Suppose these devices $\calD_k$ are ``leaky'' string-OTM's in the isolated qubits model, in the sense of Theorem \ref{thm-night-hawk}.  More precisely, suppose that for all $k \geq k_0$, 
\begin{enumerate}
\item The device $\calD_k$ stores two messages $s,t \in \set{0,1}^\ell$, where $\ell \geq k$.
\item The device $\calD_k$ uses $m$ qubits, where $k \leq m \leq k^\theta$.
\item Correctness and efficiency:  there are honest strategies for programming the device $\calD_k$ with messages $s$ and $t$, and for reading either $s$ or $t$, using only LOCC operations, and time polynomial in $k$.
\item ``Leaky'' security:  Suppose the device $\calD_k$ is programmed with two messages $(s,t)$ chosen uniformly at random.  Consider any LOCC adversary, and let $Z$ be the random variable representing the result of the adversary's measurement.  Let $M$ be any separable measurement outcome that is $\delta$-non-negligible, where $\delta = 2^{-\delta_0 k}$.  Then we have:
\begin{equation}
\label{eqn-dragonfly}
H_\infty^\eps(S,T|Z=M) \geq \alpha k,
\end{equation}
where $\eps \leq 2^{-\eps_0 k}$.
\end{enumerate}

\vskip 10pt

Now let $\calD' = \set{\calD'_k \;|\; k \geq k_0}$ be the family of devices constructed above, using $r$-wise independent random functions $F$ and $G$, with 
\begin{equation}
\label{eqn-daschund}
r = 4(\gamma+1) k^{2\theta}.
\end{equation}
(This choice of $r$ is motivated by the union bound, see equation (\ref{eqn-eagle}).  Here $\gamma$ is some universal constant, see equation (\ref{eqn-pika}).)  

\vskip 10pt

Then these devices $\calD'_k$ are ``ideal'' OTM's in the isolated qubits model, in the sense of Definition \ref{def-security}.  More precisely, for all $k \geq k_0$, with probability $\geq 1 - e^{-\Omega(k^{2\theta})}$ (over the choice of $F$ and $G$), the following statements hold:
\begin{enumerate}
\item The device $\calD'_k$ stores two messages $a_0, a_1 \in \set{0,1}$.
\item The device $\calD'_k$ uses $m$ qubits, where $k \leq m \leq k^\theta$.
\item Correctness and efficiency:  there are honest strategies for programming the device $\calD'_k$ with messages $a_0$ and $a_1$, and for reading either $a_0$ or $a_1$, using only LOCC operations, and time polynomial in $k$.
\item ``Ideal'' security:  Suppose the device $\calD'_k$ is programmed with two messages $(a_0, a_1)$ chosen uniformly at random.  Consider any LOCC adversary, and let $Z$ be the random variable representing the results of the adversary's measurements.  Then there exists a random variable $C$, which takes values in $\set{0,1}$, such that:
\begin{equation}
\begin{split}
&\norm{P_{A_C A_{1-C} C Z} - U_{A_C} \times P_{A_{1-C} C Z}}_1 \\
&\qquad\leq 4\cdot 2^{-\delta_0 k} + 2\cdot 2^{-\eps_0 k} + 2\cdot 2^{-(\alpha/8)k} + 4(r+1) \cdot 2^{-(\alpha/6)k} \\
&\qquad\leq 2^{-\Omega(k)},
\end{split}
\end{equation}
where $P_{A_C A_{1-C} C Z}$ denotes the probability density on the random variables $(A_C, A_{1-C}, C, Z)$, $P_{A_{1-C} C Z}$ denotes the marginal probability density on $(A_{1-C}, C, Z)$, and $U$ denotes the uniform distribution on $\set{0,1}$.
%
\end{enumerate}
\end{theorem}

By taking the leaky string-OTM's constructed in \cite{liu-crypto2014} (see Theorem \ref{thm-night-hawk}), and applying the above reduction, we obtain ideal OTM's in the isolated qubits model:
\begin{corollary}
There exist ideal OTM's in the isolated qubits model, in the sense of Definition \ref{def-security}.
\end{corollary}


\subsection{Overview of the proof}

We now prove Theorem \ref{thm-main}.  It is easy to see that the devices $\calD'_k$ behave correctly.  To prove that the devices $\calD'_k$ are secure, we will use a covering argument over the set of all separable measurement outcomes that can be observed by an LOCC adversary.  

We emphasize that we will be covering the set of all measurement outcomes, which are represented by POVM elements $M$, and \textit{not} the set of all LOCC adversaries, which are represented by the random variables $Z$.  To see why this is sufficient to prove security, note that for any two adversaries (represented by random variables $Z$ and $Z'$) that can observe the same measurement outcome $M$, the events $Z=M$ and $Z'=M$ are identically distributed.  

In the following argument, whenever we consider a particular measurement outcome $M$, we will also implicitly fix some adversary (represented by a random variable $Z$) that is capable of observing that outcome $M$.  We say that the scheme is ``secure at $M$'' if the scheme is secure when the adversary observes outcome $M$ (i.e., when the event $Z=M$ occurs).  

First, we will show that for any (fixed) separable measurement outcome $M$, with high probability (over the choice of the random functions $F$ and $G$ used to construct $\calD'_k$), the scheme is secure at $M$.  Next, we will construct an $\eps$-net $\Wtil$ for the set of all separable measurement outcomes, and show that with high probability (over $F$ and $G$), the scheme is secure at all points $\Mtil \in \Wtil$ simultaneously.  Finally, we will show that any separable measurement outcome $M$ can be approximated by a measurement outcome $\Mtil \in \Wtil$, such that security at $\Mtil$ implies security at $M$.  

We set the parameters in the following way:  the last part of the argument (approximating $M$ by $\Mtil \in \Wtil$) determines how small we must choose $\eps$ when constructing the $\eps$-net $\Wtil$; this determines the cardinality of $\Wtil$, which affects the union bound; this determines how large we must choose $r$ when choosing the $r$-wise independent random functions $F$ and $G$.

We now show the details:  

\textbf{Part 1:}  We begin with the following lemma, which describes what happens when we fix a particular measurement outcome $M$.  We assume that $M$ is separable and $\delta$-non-negligible; then the security guarantee for the leaky string-OTM (equation (\ref{eqn-dragonfly})) implies that:  
\begin{equation}
H_\infty^\eps(S,T|Z=M) \geq \alpha k.  
\end{equation}
The lemma introduces a random variable $C$ that indicates which of the two messages $A_0$ and $A_1$ remains unknown to the adversary; call this message $A_C$.  In addition, the lemma introduces an event $\calE$ that ``smooths'' the distribution, by excluding some low-probability failure events.  We then define a quantity $Q_c(M)$ that measures the bias of the message $A_C$, smoothed by $\calE$ and conditioned on $C=c$ and $Z=M$.  Similarly, we define a quantity $R_c(M)$ that measures the correlations between the messages $A_0$ and $A_1$, smoothed by $\calE$ and conditioned on $C=c$ and $Z=M$.  The lemma shows that, with high probability (over $F$ and $G$), $Q_c(M)$ and $R_c(M)$ are small.  
\begin{lemma}
\label{lem-singlepoint}
Fix any measurement outcome $M$ such that $H_\infty^\eps(S,T|Z=M) \geq \alpha k$.  Define $\eta = 2^{-\eta_0 k}$ where $\eta_0 = \alpha/8$.  Then there exists a random variable $C$, taking values in $\set{0,1}$, and there exists an event $\calE$, occurring with probability $\Pr(\calE|Z=M) \geq 1 - \eps - \eta$, such that the following statement holds:  Say we define, for all $c \in \set{0,1}$, 
\begin{equation}
\label{eqn-Qc}
\begin{split}
Q_c(M) &= \EE( 1_\calE \cdot (-1)^{A_C} \,|\, C=c,\, Z=M ) \\
 &= \Pr(\calE,\, A_C=0\, |\, C=c,\, Z=M) - \Pr(\calE,\, A_C=1\, |\, C=c,\, Z=M),
\end{split}
\end{equation}
which is a random variable depending on $F$ and $G$.  Then for all $c \in \set{0,1}$, and all $\lambda > 0$, 
\begin{equation}
\label{eqn-snail1}
\Pr_{FG}(\abs{Q_c(M)} \geq \lambda) \leq 2e^{1/(6r)} \sqrt{\pi r} \biggl( \frac{2^{-(\alpha/3)k} r}{e\lambda^2} \biggr)^{r/2}.
\end{equation}
In addition, say we define, for all $c \in \set{0,1}$, 
\begin{equation}
\label{eqn-Rc}
\begin{split}
R_c(M) &= \EE( 1_\calE \cdot (-1)^{A_0+A_1} \,|\, C=c,\, Z=M ),
\end{split}
\end{equation}
which is a random variable depending on $F$ and $G$.  Then for all $c \in \set{0,1}$, and all $\lambda > 0$, 
\begin{equation}
\label{eqn-snail2}
\Pr_{FG}(\abs{R_c(M)} \geq \lambda) \leq 8e^{1/(3r)} \sqrt{\pi r} \biggl( \frac{8\cdot 2^{-(\alpha/3)k} r^2}{e^2 \lambda^2} \biggr)^{r/4}.
\end{equation}
\end{lemma}

We will prove this lemma in Section \ref{sec-star1}.  This lemma is useful for the following reason:  when $Q_c(M)$ and $R_c(M)$ are small, this implies security of the devices $\calD'_k$ in the case where the adversary observes measurement outcome $M$.  We now state this observation more precisely:  

\begin{lemma}
\label{lem-hummingbird}
Fix any measurement outcome $M$, and any $c \in \set{0,1}$.  Suppose that $\abs{Q_c(M)} \leq \eps_1$ and $\abs{R_c(M)} \leq \eps_2$.  Then 
\begin{equation}
\norm{ P_{A_C A_{1-C} \calE | C=c, Z=M} - U_{A_C} \times P_{A_{1-C} \calE | C=c, Z=M} }_1 \leq \eps_1 + \eps_2,
\end{equation}
where $P_{A_C A_{1-C} \calE | C=c, Z=M}$ is the probability density 
\begin{equation}
P_{A_C A_{1-C} \calE | C=c, Z=M}(a,a') = \Pr(A_C=a,\, A_{1-C}=a',\, \calE \,|\, C=c,\, Z=M), 
\end{equation}
and $U$ denotes the uniform distribution on $\set{0,1}$.
\end{lemma}

We now prove Lemma \ref{lem-hummingbird}.  We can represent the probability density $P_{A_C A_{1-C} \calE | C=c, Z=M}$ as a vector $\vecp \in \RR^2 \tensor \RR^2$, whose entries are given by $p_{aa'} = P_{A_C A_{1-C} \calE | C=c, Z=M}(a,a')$.  Now define vectors $\vecu = \tfrac{1}{2} (1,1)$ and $\vecd = \tfrac{1}{2} (1,-1)$, which form an orthogonal basis for $\RR^2$.  Then we can write $\vecp$ in this basis:
\begin{equation}
\vecp = \alpha_{00} \vecu\tensor\vecu + \alpha_{01} \vecu\tensor\vecd + \alpha_{10} \vecd\tensor\vecu + \alpha_{11} \vecd\tensor\vecd, 
\end{equation}
for some coefficients $\alpha_{00}, \alpha_{01}, \alpha_{10}, \alpha_{11} \in \RR$.  We can write $Q_c(M)$ and $R_c(M)$ as follows:
\begin{align}
Q_c(M) &= 4(\vecd\tensor\vecu)^T\vecp = \alpha_{10}, \\
R_c(M) &= 4(\vecd\tensor\vecd)^T\vecp = \alpha_{11},
\end{align}
hence we know that $\abs{\alpha_{10}} \leq \eps_1$ and $\abs{\alpha_{11}} \leq \eps_2$.  Finally, note that the probability density $U_{A_C} \times P_{A_{1-C} \calE | C=c, Z=M}$ is represented by the following vector (call it $\vecq$):
\begin{equation}
\vecq = \vecu \tensor \bigl( 2 (\vecu^T \tensor I) \vecp \bigr) 
= \alpha_{00} \vecu\tensor\vecu + \alpha_{01} \vecu\tensor\vecd.
\end{equation}
We can combine these facts to bound the $\ell_1$ distance between $\vecp$ and $\vecq$:
\begin{equation}
\norm{\vecp-\vecq}_1 
\leq \abs{\alpha_{10}} \norm{\vecd\tensor\vecu}_1 + \abs{\alpha_{11}} \norm{\vecd\tensor\vecd}_1 
\leq \eps_1+\eps_2.
\end{equation}
This proves Lemma \ref{lem-hummingbird}.


\vskip 11pt

\textbf{Part 2:}  We let $W$ denote the set of all separable measurement outcomes, and we construct an $\eps$-net $\Wtil$ for $W$.  Specifically, we define $W$ as follows:
\begin{equation}
W = \set{M \in (\CC^{2\times 2})^{\tensor m} \;|\; M = \Tensor_{i=1}^m M_i,\, 0 \preceq M_i \preceq I}.
\end{equation}
\begin{lemma}
\label{lem-net}
For any $0 < \mu \leq 1$, there exists a set $\Wtil \subset W$, of cardinality $\abs{\Wtil} \leq (\frac{9m}{\mu})^{4m}$, which is a $\mu$-net for $W$ with respect to the operator norm $\norm{\cdot}$.
\end{lemma}

We will prove this lemma in Section \ref{sec-net}.  Now, we will use the union bound to show that, with high probability, for all $\Mtil \in \Wtil$, $Q_c(\Mtil)$ is small simultaneously.  First, we use Lemma \ref{lem-net}, and we set 
\begin{equation}
\label{eqn-sparrow}
\mu = 2^{-(\alpha/6)k} \cdot \frac{\delta^4}{4^m}
\end{equation}
(this choice is motivated by equation (\ref{eqn-finch}) below --- we choose $\mu$ small enough that the $\mu$-net gives a good approximation of any measurement outcome).  Also, recall that $k \leq m \leq k^\theta$.  Then the cardinality of $\Wtil$ is bounded by 
\begin{equation}\label{eqn-pika}
\begin{split}
\abs{\Wtil} &\leq \biggl( 9m \cdot 2^{(\alpha/6)k} \cdot \frac{4^m}{\delta^4} \biggr)^{4m} \\
&= (9m \cdot 2^{(\alpha/6)k + 4\delta_0 k + 2m})^{4m} \\
&\leq 2^{\gamma k^{2\theta}}, 
\end{split}
\end{equation}
for all sufficiently large $k$; here $\gamma$ is some universal constant.  Next, we use Lemma \ref{lem-singlepoint}, and we set 
\begin{equation}
\lambda = 2^{-(\alpha/6)k} \cdot 2r; 
\end{equation}
then we have that 
\begin{align}
\Pr_{FG}( \abs{Q_c(M)} \geq \lambda ) &\leq 2e^{1/(6r)} \sqrt{\pi r} (4er)^{-r/2}, \\
\Pr_{FG}( \abs{R_c(M)} \geq \lambda ) &\leq 8e^{1/(3r)} \sqrt{\pi r} (e^2/2)^{-r/4}.
\end{align}
Finally, we use the union bound, and we set $r$ sufficiently large (see equation (\ref{eqn-daschund})); then we have that 
\begin{equation}
\label{eqn-eagle}
\begin{split}
\Pr_{FG}\Bigl( &\exists \Mtil \in \Wtil,\, \exists c \in \set{0,1},\, 
\text{s.t. $\Mtil$ is $\delta$-non-negligible, and } 
\max\set{ \abs{Q_c(\Mtil)},\, \abs{R_c(\Mtil)} } \geq \lambda \Bigr) \\
&\leq 2\cdot 2^{\gamma k^{2\theta}} \cdot 
\Bigl( 2e^{1/(6r)} \sqrt{\pi r} (4er)^{-r/2} + 8e^{1/(3r)} \sqrt{\pi r} (e^2/2)^{-r/4} \Bigr) \\
&\leq e^{-\Omega(k^{2\theta})}.
\end{split}
\end{equation}
Hence, with high probability (over $F$ and $G$), we have that:
\begin{equation}
\label{eqn-condor}
\forall \Mtil \in \Wtil,\, \forall c \in \set{0,1},\, \text{($\Mtil$ is $\delta$-non-negligible)} 
\Rightarrow \max\set{ \abs{Q_c(\Mtil)},\, \abs{R_c(\Mtil)} } \leq \lambda.
\end{equation}
Also, note that $\lambda \leq 2^{-\Omega(k)}$.  Via Lemma \ref{lem-hummingbird}, this implies that the device $\calD'_k$ is secure in the case where the adversary observes any of the measurement outcomes in the set $\Wtil$.

\vskip 11pt

\textbf{Part 3:}  We state two lemmas that describe how an arbitrary measurement outcome $M$ can be approximated by another measurement outcome $\Mtil$.  (Implicitly, we fix some adversary that is capable of observing outcome $M$, and some other adversary that is capable of observing $\Mtil$.  We let these adversaries be represented by random variables $Z$ and $\Ztil$.)  

Roughly speaking, the first lemma shows that if $M$ is $2\delta$-non-negligible, then $\Mtil$ is $\delta$-non-negligible.
\begin{lemma}
\label{lem-non-negl}
Suppose that $M, \Mtil \in (\CC^{2\times 2})^{\tensor m}$, and $0 \preceq M \preceq I$, and $0 \preceq \Mtil \preceq I$.  Suppose that $M$ is $2\delta$-non-negligible, where $0 < \delta \leq \tfrac{1}{2}$, and $\Tr(M) \geq 1$.  Suppose that $\Mtil$ satisfies $\norm{M-\Mtil} \leq \mu$, where $\mu \leq \tfrac{2}{3} \delta \cdot 2^{-m}$.  Then $\Mtil$ is $\delta$-non-negligible.
\end{lemma}

The second lemma shows that, if the quantities $Q_c(\Mtil)$ and $R_c(\Mtil)$ are defined in terms of a random variable $\Ctil$ and an event $\calEtil$ (as in equations (\ref{eqn-Qc}) and (\ref{eqn-Rc})), then we can also define the quantities $Q_c(M)$ and $R_c(M)$ (by choosing $C$ and $\calE$ in an appropriate way), so that $Q_c(M) \approx Q_c(\Mtil)$ and $R_c(M) \approx R_c(\Mtil)$.
\begin{lemma}
\label{lem-continuity}
Suppose that $M, \Mtil \in (\CC^{2\times 2})^{\tensor m}$, and $0 \preceq M \preceq I$, and $0 \preceq \Mtil \preceq I$.  Suppose that $M$ is $2\delta$-non-negligible, where $0 < \delta \leq \tfrac{1}{2}$, and $\norm{M} = 1$.  Suppose that $\Mtil$ satisfies $\norm{M-\Mtil} \leq \mu$, where $\mu \leq \tfrac{1}{2}$, and $\Mtil$ is $\delta$-non-negligible.

Suppose there exists a random variable $\Ctil$, taking values in $\set{0,1}$, and there exists an event $\calEtil$, occurring with probability $\Pr(\calEtil|\Ztil=\Mtil)$; and let $Q_c(\Mtil)$ and $R_c(\Mtil)$ be defined in terms of $\Ctil$ and $\calEtil$, as shown in equations (\ref{eqn-Qc}) and (\ref{eqn-Rc}).

Let $0 < \tau \leq \tfrac{1}{2}$.  Then there exists a random variable $C$, taking values in $\set{0,1}$, and there exists an event $\calE$, occurring with probability $\Pr(\calE|Z=M) \geq \Pr(\calEtil|\Ztil=\Mtil) - \tau$, such that if $Q_c(M)$ and $R_c(M)$ are defined in terms of $C$ and $\calE$, then the following statements hold:
\begin{enumerate}
\item For every $c \in \set{0,1}$, either $Q_c(M) = 0$, or we have:
\begin{equation}
\abs{Q_c(M) - Q_c(\Mtil)} \leq 2\mu \biggl( \frac{2^m}{\tau\delta} \biggr)^2.
\end{equation}
\item For every $c \in \set{0,1}$, either $R_c(M) = 0$, or we have:
\begin{equation}
\abs{R_c(M) - R_c(\Mtil)} \leq 2\mu \biggl( \frac{2^m}{\tau\delta} \biggr)^2.
\end{equation}
\end{enumerate}
\end{lemma}

We will prove these two lemmas in Section \ref{sec-star2}.  Using these lemmas, we now show that the device $\calD'_k$ is secure, when the adversary observes \textit{any} separable measurement outcome $M \in W$ that is $2\delta$-non-negligible.  

Without loss of generality, suppose that $\norm{M} = 1$.  (To see this, note that without loss of generality, we can assume $M \neq 0$.  We can then multiply $M$ by a scalar factor, as long as $0 \preceq M \preceq I$, without changing the distributions of the other variables conditioned on $Z=M$.  Also note that the $\delta$-non-negligibility of $M$ is invariant under this scaling, see Definition \ref{def-delta-non-negl}.)  Note that this implies $\Tr(M) \geq 1$.

Let $\Mtil \in \Wtil$ be the nearest point in the $\mu$-net $\Wtil$; so we have $\norm{M-\Mtil} \leq \mu$, where $\mu$ is set according to equation (\ref{eqn-sparrow}).  By Lemma \ref{lem-non-negl}, $\Mtil$ is $\delta$-non-negligible.  By equation (\ref{eqn-condor}), we get that for all $c \in \set{0,1}$, $\abs{Q_c(\Mtil)} \leq \lambda$ and $\abs{R_c(\Mtil)} \leq \lambda$, where $\lambda = 2^{-(\alpha/6)k} \cdot 2r$.  

Using Lemma \ref{lem-continuity}, and setting $\tau = \delta$, we get that for every $c \in \set{0,1}$, either $Q_c(M) = 0$, or 
\begin{equation}
\label{eqn-finch}
\abs{Q_c(M) - Q_c(\Mtil)} \leq 2\mu \cdot \frac{4^m}{\delta^4} = 2\cdot 2^{-(\alpha/6)k}, 
\end{equation}
and likewise, either $R_c(M) = 0$, or $\abs{R_c(M) - R_c(\Mtil)} \leq 2\cdot 2^{-(\alpha/6)k}$.  
So we conclude that for all $c \in \set{0,1}$, 
\begin{align}
\abs{Q_c(M)} &\leq 2^{-(\alpha/6)k} \cdot 2(r+1), \\
\abs{R_c(M)} &\leq 2^{-(\alpha/6)k} \cdot 2(r+1).
\end{align}

Using Lemma \ref{lem-hummingbird}, we get that the device $\calD'_k$ is secure, for all separable $2\delta$-non-negligible measurement outcomes $M \in W$ that the adversary may observe:
\begin{equation}
\label{eqn-skunk}
\norm{ P_{A_C A_{1-C} \calE | C=c, Z=M} - U_{A_C} \times P_{A_{1-C} \calE | C=c, Z=M} }_1
\leq 2^{-(\alpha/6)k} \cdot 4(r+1) \leq 2^{-\Omega(k)}.
\end{equation}

We can write this security guarantee in a more convenient form.  Consider any LOCC adversary, and let $Z$ be the random variable representing the results of the adversary's measurements.  We can write:\footnote{There is a minor technicality involving the definition of the random variable $C$.  We have already defined $C$ whenever $Z=M$, for any $\delta$-non-negligible separable measurement outcome $M$.  We now need to define $C$ in cases where $Z=M$ and $M$ is $\delta$-negligible.  In these cases we simply define $C$ in an arbitrary way.}
\begin{equation}
\begin{split}
&\norm{P_{A_C A_{1-C} C Z} - U_{A_C} \times P_{A_{1-C} C Z}}_1 \\
&\qquad\leq \sum_M \Pr(Z=M) \, \norm{P_{A_C A_{1-C} C | Z=M} - U_{A_C} \times P_{A_{1-C} C | Z=M}}_1 \\
&\qquad\leq 4\delta + \sum_{\text{$M$ : $M$ is $2\delta$-non-negl.}} \Pr(Z=M) \, 
\norm{P_{A_C A_{1-C} C | Z=M} - U_{A_C} \times P_{A_{1-C} C | Z=M}}_1 \\
&\qquad\leq 4\delta + \sum_{\text{$M$ : $M$ is $2\delta$-non-negl.}} \Pr(Z=M) \, 
\Bigl( 2(\eps+\eta) + \norm{P_{\calE A_C A_{1-C} C | Z=M} - U_{A_C} \times P_{\calE A_{1-C} C | Z=M}}_1 \Bigr) \\
&\qquad\leq 4\delta + 2(\eps+\eta) + \sum_{\text{$M$ : $M$ is $2\delta$-non-negl.}} \Pr(Z=M) \, 
\sum_c \Pr(C=c | Z=M) \, \cdot \\
&\qquad\qquad\qquad\qquad\qquad\qquad\qquad \norm{P_{\calE A_C A_{1-C} | C=c, Z=M} - U_{A_C} \times P_{\calE A_{1-C} | C=c, Z=M}}_1 \\
&\qquad\leq 4\delta + 2(\eps+\eta) + 2^{-(\alpha/6)k} \cdot 4(r+1) \\
&\qquad\leq 4\cdot 2^{-\delta_0 k} + 2\cdot 2^{-\eps_0 k} + 2\cdot 2^{-(\alpha/8)k} + 4(r+1) \cdot 2^{-(\alpha/6)k} \\
&\qquad\leq 2^{-\Omega(k)},
\end{split}
\end{equation}
where we used the fact that $\sum_{\text{$M$ : $M$ is $2\delta$-negl.}} \Pr(Z=M) \leq 2\delta$ (see equation (\ref{eqn-delta-negl})), the fact that $\Pr(\neg\calE | Z=M) \leq \eps + \eta$ (see Lemma \ref{lem-singlepoint}), the security bound from equation (\ref{eqn-skunk}), and finally the definitions of $\delta$, $\eps$ and $\eta$ (see Theorem \ref{thm-main} and Lemma \ref{lem-singlepoint}).  This completes the proof of Theorem \ref{thm-main}.  $\square$


\subsection{Security at a single point $M$}
\label{sec-star1}

We now prove Lemma \ref{lem-singlepoint}.  We are given that $H_\infty^\eps(S,T|Z=M) \geq \alpha k$.  We will use the entropy splitting lemma (Prop. \ref{prop-rooster}).  For notational convenience, we define $\sigma_c$ to be a function that takes two arguments, and returns the first argument if $c=0$ and the second argument if $c=1$, that is, 
\begin{equation}
\sigma_c(s,t) = \begin{cases} s &\text{ if } c = 0, \\ t &\text{ if } c = 1. \end{cases}
\end{equation}
Setting $\eta = 2^{-\eta_0 k}$ and $\eta_0 = \alpha/8$, we get that there exists a random variable $C$, taking values in $\set{0,1}$, such that:
\begin{equation}
H_\infty^{\eps+\eta} (\sigma_C(S,T) \,|\, C, Z=M) \geq (\alpha/2) k - 1 - \eta_0 k \geq (\alpha/3) k.
\end{equation}

Using the definition of the smoothed conditional min-entropy, we get that there exists an event $\calE$, occurring with probability $\Pr(\calE \,|\, Z=M) \geq 1-\eps-\eta$, such that for all $c \in \set{0,1}$, and all $s \in \set{0,1}^\ell$, $\Pr(\calE, \sigma_c(S,T)=s \,|\, C=c, Z=M) \leq 2^{-(\alpha/3)k}$.  In particular, this implies that 
\begin{equation}
\label{eqn-jellyfish}
\sum_{s \in \set{0,1}^\ell} \Pr(\calE, \sigma_c(S,T)=s \,|\, C=c, Z=M)^2 \leq 2^{-(\alpha/3)k}.
\end{equation}

We now proceed to bound the quantity $Q_c(M)$.  We consider the case where $c=0$ (the $c=1$ case is similar).  In this case, we can write 
\begin{equation}
Q_0(M) = \sum_{s \in \set{0,1}^\ell} (-1)^{F(s)} \Pr(\calE, S=s \,|\, C=0, Z=M).
\end{equation}
Since $F$ is an $r$-wise independent random function, we can apply the large deviation bound in Prop. \ref{prop-kite} (making use of equation (\ref{eqn-jellyfish})).  This proves equation (\ref{eqn-snail1}).

Finally, we will bound the quantity $R_c(M)$.  We consider the case where $c=0$ (the $c=1$ case is similar).  In this case, we can write 
\begin{equation}
R_0(M) = \sum_{s,t \in \set{0,1}^\ell} (-1)^{F(s)+G(t)} \Pr(\calE, S=s, T=t \,|\, C=0, Z=M).
\end{equation}
We will bound this using Prop. \ref{prop-crayfish}.  To this end, we define a function $H:\: \set{0,1} \times \set{0,1}^\ell \rightarrow \set{0,1}$, which returns the following values:
\begin{equation}
H(i,s) = \begin{cases}
F(s) &\text{if $i=0$} \\
G(s) &\text{if $i=1$}.
\end{cases}
\end{equation}
We define a matrix $A \in \RR^{(2\cdot 2^\ell) \times (2\cdot 2^\ell)}$, whose entries are indexed by $\set{0,1} \times \set{0,1}^\ell$, and have the following values: 
\begin{equation}
A_{(i,s),(j,t)} = 
\begin{cases}
\tfrac{1}{2} \Pr(\calE, S=s, T=t \,|\, C=0, Z=M) &\text{if $(i,j)=(0,1)$} \\
\tfrac{1}{2} \Pr(\calE, S=t, T=s \,|\, C=0, Z=M) &\text{if $(i,j)=(1,0)$} \\
0 &\text{otherwise}.
\end{cases}
\end{equation}
A straightforward calculation then shows that $R_0(M)$ can be written in the form 
\begin{equation}
R_0(M) = \sum_{(i,s),(j,t)} A_{(i,s),(j,t)} \bigl( (-1)^{H(i,s)} (-1)^{H(j,t)} - \delta_{(i,s),(j,t)} \bigr).
\end{equation}

Since $F$ and $G$ are $r$-wise independent random functions, we can apply Prop. \ref{prop-crayfish}, setting $t = r/2$.\footnote{The careful reader will notice that one can actually use Prop. \ref{prop-crayfish} with $t=r$, and thereby prove a stronger bound.  The argument of Prop. \ref{prop-crayfish} still goes through, because $R_0(M)$ is \textit{bilinear} in the random variables $F(s)$ and $G(t)$, and these two groups of random variables are chosen independently of each other.}
We will use the following bounds on $\Atil$:
\begin{equation}
\begin{split}
\norm{\Atil}^2 \leq \norm{\Atil}_F^2 &= \sum_{(i,s),(j,t)} A_{(i,s),(j,t)}^2 \\
&= \tfrac{1}{2} \sum_{s,t} \Pr(\calE, S=s, T=t \,|\, C=0, Z=M)^2 \\
&\leq \tfrac{1}{2} \sum_s \biggl( \sum_t \Pr(\calE, S=s, T=t \,|\, C=0, Z=M) \biggr)^2 \\
&= \tfrac{1}{2} \sum_s \Pr(\calE, S=s \,|\, C=0, Z=M)^2 \\
&\leq \tfrac{1}{2} \cdot 2^{-(\alpha/3)k}, 
\end{split}
\end{equation}
where in the last line we used equation (\ref{eqn-jellyfish}).  We substitute into Prop. \ref{prop-crayfish}; this proves equation (\ref{eqn-snail2}).  This completes the proof of Lemma \ref{lem-singlepoint}.  $\square$


\subsection{Constructing an $\eps$-net}
\label{sec-net}

We now prove Lemma \ref{lem-net}.  First, consider the set 
\begin{equation}
V = \set{X \in \CC^{2\times 2} \;|\; \norm{X}_{\ell_\infty} \leq \sqrt{2},\, X^\dagger = X}, 
\end{equation}
where $\norm{\cdot}_{\ell_\infty}$ denotes the $\ell_\infty$ norm, viewing each $2\times 2$ matrix as a 4-dimensional vector.  Let $\delta > 0$ (we will choose a specific value for $\delta$ later).  It is easy to see that there exists a $\delta$-net $\Vtil$ for $V$, with respect to the $\ell_\infty$ norm, with cardinality $\abs{\Vtil} \leq (\tfrac{2}{\delta} + 1)^4$.  (For instance, one can describe each point in $V$ using 4 real parameters, and choose a grid with spacing $\delta\sqrt{2}$.)

Next, consider the set of single-qubit POVM elements:
\begin{equation}
U = \set{X \in \CC^{2\times 2} \;|\; 0 \preceq X \preceq I}.
\end{equation}
Note that $U \subset V$, since $\norm{X}_{\ell_\infty} \leq \norm{X}_F \leq \sqrt{2} \norm{X}$.  We will construct a $4\delta$-net $\Util$ for $U$, by ``rounding'' each point in $\Vtil$ into $U$.  Define a function $r:\: V \rightarrow U$ that maps each point in $V$ to the nearest point in $U$ with respect to the $\ell_\infty$ norm, that is, 
\begin{equation}
r(X) = \arg\min_{Y\in U} \norm{X-Y}_{\ell_\infty}.
\end{equation}
Let $\Util$ be the image of $\Vtil$ under this map, that is, $\Util = \set{r(X) \;|\; X \in \Vtil}$.  Note that $\abs{\Util} \leq \abs{\Vtil}$.  

It is easy to see that $\Util$ is a $2\delta$-net for $U$, with respect to the $\ell_\infty$ norm.  (This follows because, for any $X \in U$, there exists some $Y \in \Vtil$ such that $\norm{X-Y}_{\ell_\infty} \leq \delta$, and we know that $r(Y) \in \Util$ and $\norm{Y-r(Y)}_{\ell_\infty} \leq \delta$.)  This implies that $\Util$ is a $4\delta$-net for $U$, with respect to the operator norm $\norm{\cdot}$.  (This follows because $\norm{X} \leq \norm{X}_F \leq 2 \norm{X}_{\ell_\infty}$.)

We are now ready to consider the set $W$.  We can write $W$ in the form 
\begin{equation}
W = \set{M \;|\; M = \Tensor_{i=1}^m M_i,\, M_i \in U}.
\end{equation}
We then define $\Wtil = \set{M \;|\; M = \Tensor_{i=1}^m M_i,\, M_i \in \Util}$.  Note that $\Wtil$ has cardinality $\abs{\Wtil} \leq \abs{\Util}^m$.  

We claim that $\Wtil$ is a $4m\delta$-net for $W$, with respect to the operator norm $\norm{\cdot}$.  To see this, consider any $M \in W$, and construct some $\Mtil \in \Wtil$ that approximates it, as follows.  $M$ can be written in the form $M = \Tensor_{i=1}^m M_i$.  For each $M_i$, there is a point $\Mtil_i \in \Util$ within distance $\norm{M_i - \Mtil_i} \leq 4\delta$.  We then let $\Mtil = \Tensor_{i=1}^m \Mtil_i$.  

We upper-bound the distance $\norm{M-\Mtil}$ as follows, by defining a sequence of intermediate steps, and using the triangle inequality.  For all $s = 0,1,2,\ldots,m$, define $M^{(s)} = (\Mtil_1 \tensor \cdots \tensor \Mtil_s) \tensor (M_{s+1} \tensor \cdots \tensor M_m)$.  Then we have that $M = M^{(0)}$, $\Mtil = M^{(m)}$, and 
\begin{equation}
\begin{split}
\norm{M-\Mtil}
 &\leq \sum_{s=0}^{m-1} \norm{M^{(s)} - M^{(s+1)}} \\
 &= \sum_{s=0}^{m-1} \bigl\lVert (\Mtil_1 \tensor \cdots \tensor \Mtil_s) \tensor (M_{s+1} - \Mtil_{s+1}) \tensor (M_{s+2} \tensor \cdots \tensor M_m) \bigr\rVert \\
 &\leq 4m\delta,
\end{split}
\end{equation}
where we used the fact that $\norm{A \tensor B} = \norm{A} \, \norm{B}$.

Finally, we set $\delta = \mu/(4m)$.  Then $\Wtil$ is a $\mu$-net for $W$, with respect to the operator norm $\norm{\cdot}$.  The cardinality of $\Wtil$ is $\abs{\Wtil} \leq (\tfrac{2}{\delta} + 1)^{4m} = (\tfrac{8m}{\mu} + 1)^{4m} \leq (\tfrac{9m}{\mu})^{4m}$, provided that $\mu \leq 1$.  This proves Lemma \ref{lem-net}.  $\square$


\subsection{Continuity arguments}
\label{sec-star2}

We now prove Lemma \ref{lem-non-negl}.  Since $M$ is $2\delta$-non-negligible (with respect to some quantum state $\rho$), we have $\Pr(M) = \Tr(M\rho) \geq 2\delta \cdot 2^{-m} \Tr(M)$.  Since $\norm{M-\Mtil} \leq \mu$, and $\Tr(M) \geq 1$, we can write 
\begin{equation}
\begin{split}
\Pr(\Mtil) = \Tr(\Mtil\rho) &\geq \Tr(M\rho) - \mu \\
 &\geq 2\delta \cdot 2^{-m} \Tr(M) - \mu \\
 &\geq \delta \cdot 2^{-m} \Tr(M) + \delta \cdot 2^{-m} - \mu \\
 &\geq \delta \cdot 2^{-m} \Tr(\Mtil) - \delta \cdot \mu + \delta \cdot 2^{-m} - \mu \\
 &= \delta \cdot 2^{-m} \Tr(\Mtil) + \delta \cdot 2^{-m} - (1+\delta) \mu.
\end{split}
\end{equation}
Since $\mu \leq \tfrac{2}{3} \delta \cdot 2^{-m}$, and $\delta \leq \tfrac{1}{2}$, we have $(1+\delta) \mu \leq \delta \cdot 2^{-m}$.  By plugging into the above equation, we see that $\Mtil$ is $\delta$-non-negligible.  This proves Lemma \ref{lem-non-negl}.  $\square$

\vskip 11pt

We now prove Lemma \ref{lem-continuity}.  By assumption, there is a random variable $\Ctil$, which is defined by the probabilities $\Pr(\Ctil=c \;|\; \Ztil=\Mtil,\, S=s,\, T=t)$; and there is an event $\calEtil$, which is defined by the probabilities $\Pr(\calEtil \;|\; \Ctil=c,\, \Ztil=\Mtil,\, S=s,\, T=t)$.  Also, let $\rho_{st}$ be the quantum state used to encode messages $(s,t)$, i.e., this is the state of the one-time memory, conditioned on $S=s$ and $T=t$.  

We start by writing $Q_c(\Mtil)$ and $R_c(\Mtil)$ in a more explicit form.  First consider $Q_0(\Mtil)$, and note that $A_0 = F(S)$.  We can write $Q_0(\Mtil)$ in the form: 
\begin{equation}
\begin{split}
Q_0(\Mtil) &= \frac{1}{\Pr(\Ctil=0,\, \Ztil=\Mtil)} \sum_{s,t \in \set{0,1}^\ell} (-1)^{F(s)} 
\Pr(\calEtil,\, S=s,\, T=t,\, \Ctil=0,\, \Ztil=\Mtil) \\
 &= \frac{1}{\Pr(\Ctil=0,\, \Ztil=\Mtil)} \sum_{s,t \in \set{0,1}^\ell} (-1)^{F(s)} 
\Pr(\calEtil,\, \Ctil=0 \,|\, \Ztil=\Mtil,\, S=s,\, T=t) \Tr(\Mtil\rho_{st}) \, 4^{-\ell} \\
 &= \frac{1}{\Pr(\Ctil=0,\, \Ztil=\Mtil)} \, \Tr(\Mtil\nu_0), 
\end{split}
\end{equation}
where we define the matrix $\nu_0 \in (\CC^{2\times 2})^{\tensor m}$ as follows:
\begin{equation}
\begin{split}
\nu_0 = 4^{-\ell} \sum_{s,t \in \set{0,1}^\ell} (-1)^{F(s)} 
 &\Pr(\calEtil,\, \Ctil=0 \,|\, \Ztil=\Mtil,\, S=s,\, T=t) \, \rho_{st}.
\end{split}
\end{equation}
Note that $\norm{\nu_0}_{\Tr} \leq 1$.  In addition, we can write $\Pr(\Ctil=0,\, \Ztil=\Mtil)$ in the form: 
\begin{equation}
\begin{split}
\Pr(\Ctil=0,\, \Ztil=\Mtil) &= \sum_{s,t \in \set{0,1}^\ell} \Pr(\Ctil=0,\, \Ztil=\Mtil,\, S=s,\, T=t) \\
 &= \sum_{s,t \in \set{0,1}^\ell} \Pr(\Ctil=0 \,|\, \Ztil=\Mtil,\, S=s,\, T=t) \Tr(\Mtil\rho_{st}) \, 4^{-\ell} \\
 &= \Tr(\Mtil\xi_0), 
\end{split}
\end{equation}
where we define the matrix $\xi_0 \in (\CC^{2\times 2})^{\tensor m}$ as follows:
\begin{equation}
\xi_0 = 4^{-\ell} \sum_{s,t \in \set{0,1}^\ell} \Pr(\Ctil=0 \;|\; \Ztil=\Mtil,\, S=s,\, T=t) \, \rho_{st}.
\end{equation}
Also, note that $\norm{\xi_0}_{\Tr} \leq 1$.  We can also write similar expressions for $Q_1(\Mtil)$, $R_0(\Mtil)$ and $R_1(\Mtil)$.  We can summarize this as follows:
\begin{equation}
\label{eqn-butterfly}
Q_c(\Mtil) = \frac{\Tr(\Mtil\nu_c)}{\Tr(\Mtil\xi_c)}, \qquad
R_c(\Mtil) = \frac{\Tr(\Mtil\theta_c)}{\Tr(\Mtil\xi_c)}, 
\end{equation}
where $\nu_c, \theta_c, \xi_c \in (\CC^{2\times 2})^{\tensor m}$ satisfy $\norm{\nu_c}_{\Tr} \leq 1$, $\norm{\theta_c}_{\Tr} \leq 1$ and $\norm{\xi_c}_{\Tr} \leq 1$.  

We now consider the measurement outcome $M$.  We will construct a random variable $C$ and an event $\calE$, which will allow us to define the quantities $Q_c(M)$ and $R_c(M)$.  Roughly speaking, $C$ and $\calE$ (conditioned on $Z=M$) will behave similarly to $\Ctil$ and $\calEtil$ (conditioned on $\Ztil=\Mtil$).  However, if there exists some $c \in \set{0,1}$ for which the probability $\Pr(C=c\, |\, Z=M)$ is unusually small, then we will define $\calE$ to exclude this event, in order to avoid situations where $Q_c(M)$ ``blows up'' because the denominator is very small.

Formally, we construct the random variable $C$ and the event $\calE$ by specifying the following probabilities (for all $c \in \set{0,1}$ and $s,t \in \set{0,1}^\ell$):
\begin{align}
\Pr(C=c\, |\, Z=M,\, S=s,\, T=t) &= \Pr(\Ctil=c\, |\, \Ztil=\Mtil,\, S=s,\, T=t), \\
\Pr(\calE\, |\, C=c,\, Z=M,\, S=s,\, T=t) &= 
\begin{cases}
0 \quad\text{ if } \Pr(C=c\, |\, Z=M) < \tau, \\
\Pr(\calEtil\, |\, \Ctil=c,\, \Ztil=\Mtil,\, S=s,\, T=t) \quad\text{ otherwise}.
\end{cases}
\end{align}

We now show that $\Pr(\calE\, |\, Z=M) \geq \Pr(\calEtil\, |\, \Ztil=\Mtil) - \tau$.  Let us say that $c \in \set{0,1}$ is ``bad'' if $\Pr(C=c\, |\, Z=M) < \tau$.  There are two possible values, 0 and 1, and at most one of them can be bad.  If neither one is bad, then $\Pr(\calE\, |\, Z=M) = \Pr(\calEtil\, |\, \Ztil=\Mtil)$.  If one particular value (say 0) is bad, then we have:
\begin{equation}
\begin{split}
\Pr(\calE \,|\, Z=M) &\geq \Pr(\calE \,|\, C=1,\, Z=M) \Pr(C=1 \,|\, Z=M) \\
 &= \Pr(\calEtil \,|\, \Ctil=1,\, \Ztil=\Mtil) \Pr(\Ctil=1 \,|\, \Ztil=\Mtil) \\
 &= \Pr(\calEtil \,|\, \Ztil=\Mtil) - \Pr(\calEtil \,|\, \Ctil=0,\, \Ztil=\Mtil) \Pr(\Ctil=0 \,|\, \Ztil=\Mtil) \\
 &> \Pr(\calEtil \,|\, \Ztil=\Mtil) - \tau.
\end{split}
\end{equation}

We now define $Q_c(M)$ in terms of $C$ and $\calE$, using equation (\ref{eqn-Qc}).  
Note that if $c$ is bad, then $\Pr(\calE \,|\, C=c,\, Z=M) = 0$, which implies $Q_c(M) = 0$.  

We will show that if $c$ is not bad, then $Q_c(M) \approx Q_c(\Mtil)$.  When $c$ is not bad, the events $C=c$ and $\calE$ (conditioned on the events $Z=M$, $S=s$ and $T=t$) have exactly the same probabilities as the events $\Ctil=c$ and $\calEtil$ (conditioned on the events $\Ztil=\Mtil$, $S=s$ and $T=t$).  So we can write $Q_c(M)$ in the form 
\begin{equation}
Q_c(M) = \frac{\Tr(M\nu_c)}{\Tr(M\xi_c)},
\end{equation}
where $\nu_c$ and $\xi_c$ are the \textit{same} matrices used to express $Q_c(\Mtil)$ in equation (\ref{eqn-butterfly}).  In addition, we can lower-bound $\Tr(M\xi_c)$ and $\Tr(\Mtil\xi_c)$ as follows:
\begin{align}
\Tr(M\xi_c) &= \Pr(C=c,\, M) \geq \tau \Pr(M) \\
 &\geq \tau \cdot 2\delta \cdot 2^{-m} \Tr(M) \geq \tau \cdot 2\delta \cdot 2^{-m} \norm{M} \\
 &\geq \tau \cdot 2\delta \cdot 2^{-m}, \\
\Tr(\Mtil\xi_c) &= \Pr(\Ctil=c,\, \Mtil) \geq \tau \Pr(\Mtil) \\
 &\geq \tau \delta \cdot 2^{-m} \Tr(\Mtil) \geq \tau \delta \cdot 2^{-m} \norm{\Mtil} \\
 &\geq \tau \delta \cdot 2^{-m} (1-\mu) \geq \tau \delta \cdot 2^{-m} \cdot \tfrac{1}{2}.
\end{align}
Now we can write $Q_c(M) - Q_c(\Mtil)$ as follows:
\begin{equation}
Q_c(M) - Q_c(\Mtil) = \frac{\Tr((M-\Mtil)\nu_c)}{\Tr(M\xi_c)} + \Tr(\Mtil\nu_c) \frac{\Tr((\Mtil-M)\xi_c)}{\Tr(M\xi_c)\Tr(\Mtil\xi_c)}.
\end{equation}
We can then upper-bound this quantity:
\begin{equation}
\begin{split}
\abs{Q_c(M) - Q_c(\Mtil)}
 &\leq \frac{\mu}{\tau \cdot 2\delta \cdot 2^{-m}} + (1+\mu) \frac{\mu}{\tau \cdot 2\delta \cdot 2^{-m} \cdot \tau \delta \cdot 2^{-m} \cdot \tfrac{1}{2}} \\
 &= \frac{\mu}{\tau \cdot 2\delta \cdot 2^{-m}} \biggl( 1 + \frac{(1+\mu)}{\tau \delta \cdot 2^{-m} \cdot \tfrac{1}{2}} \biggr) \\
 &\leq 2\mu \biggl( \frac{2^m}{\tau\delta} \biggr)^2.
\end{split}
\end{equation}

Similarly, we define $R_c(M)$ in terms of $C$ and $\calE$, using equation (\ref{eqn-Rc}).  Using the same argument as above, we see that if $c$ is bad, then $R_c(M) = 0$, and if $c$ is not bad, then $R_c(M) \approx R_c(\Mtil)$.  This completes the proof of Lemma \ref{lem-continuity}.  $\square$


\section{Beyond the isolated qubits model}

We now describe a class of adversaries who can perform a polynomial number of 2-qubit entangling operations, in addition to unbounded LOCC.  In particular, we will choose some ``depth'' parameter $d$ (which may grow polynomially with the security parameter $k$), and we will consider adversaries who can apply quantum circuits whose depth is bounded by $d$.  These kinds of attacks may be feasible in real physical systems, where one can perform noisy entangling gates.  Intuitively, one may expect that the noise will accumulate when the adversary applies a long sequence of entangling gates; so it is easier for the adversary to apply shallow (low-depth) quantum circuits.

We will then show that our privacy amplification result for one-time memories (Theorem \ref{thm-main}) still holds against these depth-$d$ adversaries, where $d$ can grow polynomially in $k$, and the privacy amplification technique now runs in time polynomial in $d$.  

First, we will need a few definitions.  Let $\calE:\: \rho \mapsto \sum_i K_i \rho K_i^\dagger$ be a generalized quantum measurement.  We say that $\calE$ is \textit{2-local} if every Kraus operator $K_i$ can be written as a tensor product of 2-qubit operators (where different Kraus operators $K_i$ may arrange the qubits into pairs in different ways).  As a simple example, if $\calE_1, \calE_2, \ldots, \calE_\ell$ are 2-qubit quantum measurements, then $\calE_1 \tensor \calE_2 \tensor \cdots \tensor \calE_\ell$ is a 2-local quantum measurement on $2\ell$ qubits.  

Note that 2-local measurements can be viewed as a generalization of separable measurements, in the following sense.  First, if $\calE$ is separable, then $\calE$ is 2-local.  Also, if $\calE_1$ and $\calE_2$ are separable, and $\calF$ is 2-local, then $\calE_2 \circ \calF \circ \calE_1$ is 2-local.  Thus any 2-local measurement can include a separable measurement (and in particular, an LOCC measurement) ``for free.''

We say that an adversary is \textit{2-local with depth $d$} if it performs a measurement of the form $\calE = \calE_d \circ \calE_{d-1} \circ \cdots \circ \calE_1$, where $\calE_1, \calE_2, \ldots, \calE_d$ are 2-local measurements.  That is, the adversary first performs the measurement $\calE_1$, obtains a classical measurement outcome $i_1$, then performs the measurement $\calE_2$, obtains a classical measurement outcome $i_2$, and so on; after the final measurement $\calE_d$, the post-measurement quantum state is discarded.  

We say that the corresponding POVM element $M_{i_1,i_2,\ldots,i_d}$ is \textit{2-local with depth $d$}.  We can write it in the following form:
\begin{equation}
M_{i_1,i_2,\ldots,i_d} = ( K_{1,i_1}^\dagger K_{2,i_2}^\dagger \cdots K_{d,i_d}^\dagger ) \, ( K_{d,i_d} \cdots K_{2,i_2} K_{1,i_1} ), 
\end{equation}
where the $K_{a,i_a}$ denote the Kraus operators of the measurement $\calE_a$, that is, $\calE_a(\rho) = \sum_{i_a} K_{a,i_a} \rho K_{a,i_a}^\dagger$, and each $K_{a,i_a}$ can be written as a tensor product of 2-qubit operators.  

We now extend our privacy amplification result for one-time memories (Theorem \ref{thm-main}) to the case of 2-local depth-$d$ adversaries.  
\begin{theorem}
\label{thm-main2}
Fix some constant $\varphi \geq 0$.  

Suppose that $\calD$ is a family of ``leaky'' string-OTM's, as described in Theorem \ref{thm-main}, but with a stronger security guarantee, which holds for all measurement outcomes that are 2-local with depth $d \leq k^\varphi$ (rather than for all separable measurement outcomes).  

Now construct a new family of devices $\calD'$, as described in Theorem \ref{thm-main}, but where we set the parameter $r$ (for the $r$-wise independent random functions $F$ and $G$) as follows:
\begin{equation}
\label{eqn-daschund2}
r = 4(\gamma+1) k^{2\theta+\varphi}.
\end{equation}

Then these devices $\calD'$ are ``ideal'' OTM's, as described in Theorem \ref{thm-main}, but again with a stronger security guarantee, which holds for all measurement outcomes that are 2-local with depth $d \leq k^\varphi$ (rather than for all separable measurement outcomes).  
\end{theorem}

Thus, if one could construct ``leaky'' string-OTM's that were secure against 2-local depth-$d$ adversaries, then one would immediately get ``ideal'' bit-OTM's in this setting.  Unfortunately, the leaky string-OTM's from \cite{liu-crypto2014} are not known to be secure in this setting, and so we leave this as an open problem.


\subsection{Overview of the proof}

We prove Theorem \ref{thm-main2} using the same approach as for Theorem \ref{thm-main}.  Most of the argument is unchanged; the key difference is in Lemma \ref{lem-net}, where we now want to construct an $\eps$-net for the set of all 2-local depth-$d$ measurement outcomes (rather than the set of all separable measurement outcomes).  

Let $\Lambda$ be the set of all 2-local depth-$d$ measurement outcomes:
\begin{equation}
\Lambda = \set{ M \in (\CC^{2\times 2})^{\tensor m} \;|\; M = (K_1^\dagger \cdots K_d^\dagger)\, (K_d \cdots K_1),\, \text{where } K_1,\ldots, K_d \in L}, 
\end{equation}
where $L$ is the set of all operators $K \in (\CC^{2\times 2})^{\tensor m}$ that can be written as tensor products of 2-qubit operators having operator norm at most 1.  We will construct an $\eps$-net for $\Lambda$, using the following lemma:
\begin{lemma}
\label{lem-net2}
For any $0 < \mu \leq 1$, there exists a set $\Lambdatil \subset \Lambda$, of cardinality $\abs{\Lambdatil} \leq \bigl( \frac{24dm^{17/16}}{\mu} \bigr)^{16md}$, which is a $\mu$-net for $\Lambda$ with respect to the operator norm $\norm{\cdot}$.
\end{lemma}

We will prove this lemma in Section \ref{sec-star3}.  Now, we set 
\begin{equation}
\label{eqn-sparrow2}
\mu = 2^{-(\alpha/6)k} \cdot \frac{\delta^4}{4^m}
\end{equation}
(the same as in the proof of Theorem \ref{thm-main}).  Also, recall that $k \leq m \leq k^\theta$, and $d \leq k^\varphi$.  Then the cardinality of $\Lambdatil$ is bounded by 
\begin{equation}\label{eqn-pika2}
\begin{split}
\abs{\Lambdatil} 
&\leq \biggl( 24dm^{17/16} \cdot 2^{(\alpha/6)k} \cdot \frac{4^m}{\delta^4} \biggr)^{16md} \\
&= \bigl( 24dm^{17/16} \cdot 2^{(\alpha/6)k + 4\delta_0 k + 2m} \bigr)^{16md} \\
&\leq 2^{\gamma k^{2\theta+\varphi}}, 
\end{split}
\end{equation}
for all sufficiently large $k$; here $\gamma$ is some universal constant.  This bound plays the role of equation (\ref{eqn-pika}) in the proof of Theorem \ref{thm-main}.  

One then continues with the same argument as in Theorem \ref{thm-main}:  one uses the union bound over the set $\Lambdatil$, while setting the parameter $r$ sufficiently large (see equation (\ref{eqn-daschund2})).  This gives a result similar to equation (\ref{eqn-eagle}).  

The rest of the proof is the same as for Theorem \ref{thm-main}.  This completes the proof of Theorem \ref{thm-main2}.  $\square$


\subsection{Constructing an $\eps$-net}
\label{sec-star3}

We now prove Lemma \ref{lem-net2}.  First, consider the set 
\begin{equation}
V = \set{X \in \CC^{4\times 4} \;|\; \norm{X}_{\ell_\infty} \leq 2}.
\end{equation}
Let $\delta > 0$ (we will choose a specific value for $\delta$ later).  It is easy to see that there exists a $\delta$-net $\Vtil$ for $V$, with respect to the $\ell_\infty$ norm, with cardinality $\abs{\Vtil} \leq (\tfrac{2\sqrt{2}}{\delta} + 1)^{32}$.  (For instance, one can describe each point in $V$ with 32 real parameters, and choose a grid with spacing $\delta\sqrt{2}$.)

Next, consider the set of 2-qubit Kraus operators:
\begin{equation}
U = \set{X \in \CC^{4\times 4} \;|\; \norm{X} \leq 1}.
\end{equation}
Note that $U \subset V$, since $\norm{X}_{\ell_\infty} \leq \norm{X}_F \leq 2 \norm{X}$.  We will construct an $8\delta$-net $\Util$ for $U$, by taking the points in $\Vtil$ and ``rounding'' them into $U$.  Define a function $r:\: V \rightarrow U$ that maps each point in $V$ to the nearest point in $U$ with respect to the $\ell_\infty$ norm, that is, 
\begin{equation}
r(X) = \arg\min_{Y\in U} \norm{X-Y}_{\ell_\infty}.
\end{equation}
Let $\Util$ be the image of $\Vtil$ under this map, that is, $\Util = \set{r(X) \;|\; X \in \Vtil}$.  Note that $\abs{\Util} \leq \abs{\Vtil}$.  

It is easy to see that $\Util$ is a $2\delta$-net for $U$, with respect to the $\ell_\infty$ norm.  (This follows because, for any $X \in U$, there exists some $Y \in \Vtil$ such that $\norm{X-Y}_{\ell_\infty} \leq \delta$, and we know that $r(Y) \in \Util$ and $\norm{Y-r(Y)}_{\ell_\infty} \leq \delta$.)  This implies that $\Util$ is an $8\delta$-net for $U$, with respect to the operator norm $\norm{\cdot}$.  (This follows because $\norm{X} \leq \norm{X}_F \leq 4 \norm{X}_{\ell_\infty}$.)

Next, we let $L$ be the set of all operators $K \in (\CC^{2\times 2})^{\tensor m}$ that can be written as tensor products of 2-qubit operators in $U$.  
We then define $\Ltil$ to be the set of all operators $K \in (\CC^{2\times 2})^{\tensor m}$ that can be written as tensor products of 2-qubit operators in $\Util$.  Note that $\Ltil$ has cardinality $\abs{\Ltil} \leq m! \, \abs{\Util}^{m/2}$, since every operator $K \in \Ltil$ can be written in the form $\Tensor_{j=1}^{m/2} K_j$ (where $K_j \in \Util$) conjugated with a permutation of the qubits.  (For simplicity, let us assume that $m$ is even.)

We claim that $\Ltil$ is a $4m\delta$-net for $L$, with respect to the operator norm $\norm{\cdot}$.  To see this, consider any $K \in L$, and construct some $\Ktil \in \Ltil$ that approximates it as follows.  First, relabel the qubits so that $K$ can be written in the form $K = \Tensor_{j=1}^{m/2} K_j$ (where $K_j \in U$).  For each $K_j$, there is a point $\Ktil_j \in \Util$ within distance $\norm{K_j - \Ktil_j} \leq 8\delta$.  We then define $\Ktil = \Tensor_{j=1}^{m/2} \Ktil_j$.  

We upper-bound the distance $\norm{K-\Ktil}$ as follows, by defining a sequence of intermediate steps, and using the triangle inequality.  For all $s = 0,1,2,\ldots,m/2$, define $K^{(s)} = (\Ktil_1 \tensor \cdots \tensor \Ktil_s) \tensor (K_{s+1} \tensor \cdots \tensor K_{m/2})$.  Then we have that $K = K^{(0)}$, $\Ktil = K^{(m/2)}$, and 
\begin{equation}
\begin{split}
\norm{K-\Ktil} &\leq \sum_{s=0}^{m/2-1} \norm{K^{(s)} - K^{(s+1)}} \\
 &= \sum_{s=0}^{m/2-1} \bigl\lVert (\Ktil_1 \tensor \cdots \tensor \Ktil_s) \tensor (K_{s+1} - \Ktil_{s+1}) \tensor (K_{s+2} \tensor \cdots \tensor K_{m/2}) \bigr\rVert \\
 &\leq (m/2) \, 8\delta = 4m\delta,
\end{split}
\end{equation}
where we used the fact that $\norm{A \tensor B} = \norm{A} \, \norm{B}$.

Finally, we consider the set $\Lambda$ of all 2-local depth-$d$ measurement outcomes:
\begin{equation}
\Lambda = \set{ M \in (\CC^{2\times 2})^{\tensor m} \;|\; M = (K_1^\dagger \cdots K_d^\dagger)\, (K_d \cdots K_1),\, \text{where } K_1,\ldots, K_d \in L}.
\end{equation}
We then define the set $\Lambdatil$ as follows:
\begin{equation}
\Lambdatil = \set{ M \in (\CC^{2\times 2})^{\tensor m} \;|\; M = (K_1^\dagger \cdots K_d^\dagger)\, (K_d \cdots K_1),\, \text{where } K_1,\ldots, K_d \in \Ltil}.
\end{equation}
Note that $\Lambdatil$ has cardinality $\abs{\Lambdatil} \leq \abs{\Ltil}^d$.

We claim that $\Lambdatil$ is an $8dm\delta$-net for $\Lambda$, with respect to the operator norm $\norm{\cdot}$.  To see this, consider any $M \in \Lambda$, and construct some $\Mtil \in \Lambdatil$ that approximates it as follows.  $M$ can be written in the form $M = (K_1^\dagger \cdots K_d^\dagger)\, (K_d \cdots K_1)$ (where $K_j \in L$).  For each $K_j$, there is a point $\Ktil_j \in \Ltil$ within distance $\norm{K_j - \Ktil_j} \leq 4m\delta$.  We then let $\Mtil = (\Ktil_1^\dagger \cdots \Ktil_d^\dagger)\, (\Ktil_d \cdots \Ktil_1)$.  

We upper-bound the distance $\norm{M-\Mtil}$ as follows, by defining a sequence of intermediate steps, and using the triangle inequality.  For all $s = 0,1,2,\ldots,2d$, define $M^{(s)}$ to be an operator of the form $(K_1^\dagger \cdots K_d^\dagger)\, (K_d \cdots K_1)$, where the first $s$ factors (reading from left to right) have tilde's, and the remaining $2d-s$ factors do not have tilde's.  Then we have that $M = M^{(0)}$, $\Mtil = M^{(2d)}$, and 
\begin{equation}
\begin{split}
\norm{M-\Mtil} &\leq \sum_{s=0}^{2d-1} \norm{M^{(s)} - M^{(s+1)}} \\
 &= \sum_{s=0}^{d-1} \bigl\lVert (\Ktil_1^\dagger \cdots \Ktil_s^\dagger) (K_{s+1}^\dagger - \Ktil_{s+1}^\dagger) (K_{s+2}^\dagger \cdots K_d^\dagger) \, (K_d \cdots K_1) \bigr\rVert \\
 &+ \sum_{s=d}^{2d-1} \bigl\lVert (\Ktil_1^\dagger \cdots \Ktil_d^\dagger) \, (\Ktil_d \cdots \Ktil_{2d-s+1}) (K_{2d-s} - \Ktil_{2d-s}) (K_{2d-s-1} \cdots K_1) \bigr\rVert \\
 &\leq 2d \cdot 4m\delta = 8dm\delta,
\end{split}
\end{equation}
where we used the fact that $\norm{AB} \leq \norm{A} \, \norm{B}$.

Finally, we set $\delta = \mu/(8dm)$.  Then $\Lambdatil$ is a $\mu$-net for $\Lambda$, with respect to the operator norm $\norm{\cdot}$.  The cardinality of $\Lambdatil$ is 
\begin{equation}
\begin{split}
\abs{\Lambdatil} &\leq \abs{\Ltil}^d \leq (m!)^d \, \abs{\Util}^{md/2} \\
 &\leq (m!)^d \, \bigl( \tfrac{2\sqrt{2}}{\delta}+1 \bigr)^{16md} \\
 &\leq m^{md} \, \bigl( \tfrac{16\sqrt{2}dm}{\mu} + 1 \bigr)^{16md} \\
 &\leq \bigl( \tfrac{24dm^{17/16}}{\mu} \bigr)^{16md}, 
\end{split}
\end{equation}
provided that $\mu \leq 1$.  This proves Lemma \ref{lem-net2}.  $\square$


\subsection*{Acknowledgements}

It is a pleasure to thank Fang Song, Hong-Sheng Zhou, Christian Schaffner, Marc Kaplan, Frederic Dupuis, Stephen Jordan, and Ray Perlner, for helpful discussions and suggestions.  Special thanks to the anonymous referees of Eurocrypt, for pointing out a weakness in our security definition (which we have now fixed), and for several suggestions to improve the presentation of these results.  This work is a contribution of NIST, an agency of the US government, and is not subject to US copyright.



\appendix

\section{Large deviation bounds}

\subsection{Sums of several random variables}
\label{app-starfish}

We prove Proposition \ref{prop-kite}, a large-deviation bound for sums of $t$-wise independent random variables.  The proof is essentially the same as in \cite{bellare-rompel}, with minor modifications because the random variables have different distributions.  First we use Markov's inequality:
\begin{equation}
\Pr(\abs{Y} \geq \lambda) \leq \frac{1}{\lambda^t} \EE(\abs{Y}^t) = \frac{1}{\lambda^t} \EE(Y^t).
\end{equation}

Now let $\Gamma_1,\ldots,\Gamma_N$ be independent random variables uniformly distributed in $\set{0,1}$, and define $\Ytil = \sum_{x=1}^N (-1)^{\Gamma_x} a_x$ (the same expression as $Y$, but replacing the random variables $H(x)$ with $\Gamma_x$).  Since the $H(x)$ are $t$-wise independent, we know that 
\begin{equation}
\EE(Y^t) = \EE(\Ytil^t).
\end{equation}
(To see this, write $Y = \sum_{x=1}^N (1-2H(x)) a_x$, which is a linear function of the random variables $H(x)$; hence $Y^t$ is a degree-$t$ polynomial in the variables $H(x)$.)

We can then bound $\EE(\Ytil^t)$ as follows (using Hoeffding's inequality, letting $y = x^{2/t}/2v$, and using Stirling's inequality):
\begin{equation}
\begin{split}
\EE(\Ytil^t) &= \int_0^\infty \Pr(\Ytil^t > x) dx = \int_0^\infty \Pr(|\Ytil| > x^{1/t}) dx \\
 &\leq \int_0^\infty 2\exp(-x^{2/t}/2v) dx \\
 &= 2 \int_0^\infty e^{-y} (2v)^{t/2} (t/2) y^{(t/2)-1} dy \\
 &= 2(2v)^{t/2} \cdot (t/2)! \\
 &< 2(2v)^{t/2} \cdot e^{1/(6t)} \sqrt{\pi t} (t/2e)^{t/2} \\
 &= 2e^{1/(6t)} \sqrt{\pi t} (vt/e)^{t/2}.
\end{split}
\end{equation}
This proves Proposition \ref{prop-kite}.  $\square$


\subsection{Quadratic functions of several random variables}
\label{app-crayfish}

We prove Proposition \ref{prop-crayfish}, a large-deviation bound for quadratic functions of $2t$-wise independent random variables.  We begin by re-stating the Hanson-Wright inequality \cite{hanson-wright}, specialized to the case of Rademacher random variables, and with explicit constants.  (Note that a stronger bound is possible, involving $A$ in place of $\Atil$; see \cite{hanson-wright-RV} for details.)
\begin{lemma}
\label{lem-hw}
Let $\xi_1,\ldots,\xi_N$ be independent random variables, uniformly distributed in $\set{1,-1}$.  Let $A \in \RR^{N\times N}$ be a symmetric matrix, $A^T = A$.  Define the random variable 
\begin{equation}
T = \sum_{i,j=1}^N A_{ij} ( \xi_i\xi_j - \EE \xi_i\xi_j ).
\end{equation}

Then $\EE T = 0$, and we have the following large-deviation bound:  for any $\lambda > 0$, 
\begin{equation}
\Pr(T \geq \lambda) \leq \exp\biggl( -\frac{1}{8} \min\biggl\lbrace \frac{\lambda}{\norm{\Atil}},\; \frac{\lambda^2}{\norm{\Atil}_F^2} \biggr\rbrace \biggr), 
\end{equation}
where $\Atil \in \RR^{N\times N}$ is the entry-wise absolute value of $A$, that is, $\Atil_{ij} = \abs{A_{ij}}$.
\end{lemma}

We now consider the random variable $S$ in Prop. \ref{prop-crayfish}.  $S$ is identical to $T$, except that the fully-independent random variables $\xi_i$ are replaced by the $2t$-wise-independent random variables $(-1)^{H(x)}$.  We bound $S$ using the same argument as in \cite{bellare-rompel}.  First we use Markov's inequality:
\begin{equation}
\Pr(\abs{S} \geq \lambda) \leq \frac{1}{\lambda^t} \EE(\abs{S}^t) = \frac{1}{\lambda^t} \EE(S^t).
\end{equation}
Since the $H(x)$ are $2t$-wise independent, and $S$ is a quadratic function of the variables $(-1)^{H(x)} = 1-2H(x)$, we know that 
\begin{equation}
\EE(S^t) = \EE(T^t).
\end{equation}

We can then bound $\EE(T^t)$ as follows (using Lemma \ref{lem-hw}, letting $\Lambda = \norm{\Atil}$ and $\Lambda_F = \norm{\Atil}_F$, letting $y = \tfrac{1}{8} (x^{2/t}/\Lambda_F^2)$ and $z = \tfrac{1}{8} (x^{1/t}/\Lambda)$, and using Stirling's inequality):
\begin{equation}
\begin{split}
\EE(T^t) &= \int_0^\infty \Pr(T^t > x) dx = \int_0^\infty \Pr(|T| > x^{1/t}) dx \\
 &\leq \int_0^\infty 2\exp\biggl( -\frac{1}{8} \min\biggl\lbrace \frac{x^{1/t}}{\Lambda},\; \frac{x^{2/t}}{\Lambda_F^2} \biggr\rbrace \biggr) dx \\
 &= \int_0^{(\Lambda_F^2/\Lambda)^t} 2\exp\bigl( -\tfrac{1}{8} (x^{2/t}/\Lambda_F^2) \bigr) dx
  + \int_{(\Lambda_F^2/\Lambda)^t}^\infty 2\exp\bigl( -\tfrac{1}{8} (x^{1/t}/\Lambda) \bigr) dx \\
 &= \int_0^{(1/8)\Lambda_F^2/\Lambda^2} 2e^{-y} (8\Lambda_F^2)^{t/2} (t/2) y^{(t/2)-1} dy
  + \int_{(1/8)\Lambda_F^2/\Lambda^2}^\infty 2e^{-z} (8\Lambda)^t t z^{t-1} dz \\
 &< 2 (8\Lambda_F^2)^{t/2} \cdot (t/2)!
  + 2 (8\Lambda)^t \cdot t! \\
 &< 2 (8\Lambda_F^2)^{t/2} \cdot e^{1/(6t)} \sqrt{\pi t} (t/2e)^{t/2}
  + 2 (8\Lambda)^t \cdot e^{1/(12t)} \sqrt{2\pi t} (t/e)^t \\
 &= 2e^{1/(6t)} \sqrt{\pi t} \Bigl( \frac{4\Lambda_F^2 t}{e} \Bigr)^{t/2}
  + 2e^{1/(12t)} \sqrt{2\pi t} \Bigl( \frac{8\Lambda t}{e} \Bigr)^t.
\end{split}
\end{equation}
This proves Proposition \ref{prop-crayfish}.  $\square$



\begin{thebibliography}{99}

\bibitem{nogo1} H.-K. Lo and H.F. Chau, ``Is quantum bit commitment really possible?'' \textit{Phys. Rev. Lett.} 78, 3410 (1997).
\bibitem{nogo2} H.-K. Lo, ``Insecurity of quantum secure computations,'' \textit{Phys. Rev. A}, 56(2): 1154-1162 (1997).
\bibitem{nogo3} D. Mayers, ``Unconditionally secure quantum bit commitment is impossible,'' \textit{Phys. Rev. Lett.}, 78:3414-3417 (1997).
\bibitem{nogo4} H. Buhrman, M. Christandl and C. Schaffner, ``Complete Insecurity of Quantum Protocols for Classical Two-Party Computation,'' \textit{Phys. Rev. Lett.} 109, 160501 (2012).

\bibitem{liu-crypto2014} Y.-K. Liu, ``Single-Shot Security for One-Time Memories in the Isolated Qubits Model,'' \textit{CRYPTO 2014}, vol. 2, pp.19-36 (2014).
\bibitem{liu-itcs2014} Y.-K. Liu, ``Building one-time memories from isolated qubits,'' \textit{ITCS 2014}, pp.269-286.

\bibitem{saeedi-2013} K. Saeedi et al, ``Room-Temperature Quantum Bit Storage Exceeding 39 Minutes Using Ionized Donors in Silicon-28,'' \textit{Science} 342 (6160) pp.830-833 (2013).
\bibitem{dreau-2013} A. Dreau et al, ``Single-Shot Readout of Multiple Nuclear Spin Qubits in Diamond under Ambient Conditions,'' \textit{Phys. Rev. Lett.} 110, 060502 (2013).

\bibitem{GKR} S. Goldwasser, Y.T. Kalai and G.N. Rothblum, ``One-Time Programs,'' \textit{CRYPTO 2008}, pp.39-56.
\bibitem{goyal} V. Goyal, Y. Ishai, A. Sahai, R. Venkatesan and A. Wadia, ``Founding Cryptography on Tamper-Proof Hardware Tokens,'' \textit{TCC 2010}, pp.308-326.
\bibitem{bellare} M. Bellare, V. T. Hoang and P. Rogaway, ``Adaptively Secure Garbling with Applications to One-Time Programs and Secure Outsourcing,'' \textit{ASIACRYPT 2012}, pp.134-153.
\bibitem{broadbent} A. Broadbent, G. Gutoski and D. Stebila, ``Quantum one-time programs,'' \textit{CRYPTO 2013}, pp.344-360.

\bibitem{vadhan} S. P. Vadhan, ``Pseudorandomness,'' \textit{Foundations and Trends in Theoretical Computer Science}, Vol. 7, Issue 1–3 (2011), pp.1-336.
\bibitem{bellare-rompel} M. Bellare and J. Rompel, ``Randomness-Efficient Oblivious Sampling,'' \textit{FOCS 1994}, pp.276-287.
\bibitem{srinivasan} J. P. Schmidt, A. Siegel and A. Srinivasan, ``Chernoff-Hoeffding Bounds for Applications with Limited Independence,'' \textit{SIAM J. Discrete Math.} 8(2), pp.223-250 (1995).
\bibitem{hanson-wright} D. L. Hanson and F. T. Wright, ``A Bound on Tail Probabilities for Quadratic Forms in Independent Random Variables,'' \textit{Ann. Math. Stat.}, Vol. 42, No. 3, pp.1079-1083 (1971).
\bibitem{hanson-wright-RV} M. Rudelson and R. Vershynin, ``Hanson-Wright inequality and sub-gaussian concentration,'' \textit{Electronic Communications in Probability} 18 (2013), pp.1-9.

\bibitem{wiesner} S. Wiesner, ``Conjugate coding,'' \textit{ACM SIGACT News}, Volume 15, Issue 1, 1983, pp.78-88; original manuscript written circa 1970. 

\bibitem{salvail} L. Salvail, ``Quantum Bit Commitment from a Physical Assumption,'' \textit{CRYPTO 1998}, pp.338-353.
\bibitem{pastawski} F. Pastawski, N. Y. Yao, L. Jiang, M. D. Lukin and J. I. Cirac, ``Unforgeable Noise-Tolerant Quantum Tokens,'' \textit{Proc. Nat. Acad. Sci.} 109, 16079-16082 (2012).
\bibitem{all-but-one} N. J. Bouman, S. Fehr, C. Gonzalez-Guillen and C. Schaffner, ``An All-But-One Entropic Uncertainty Relation, and Application to Password-Based Identification,'' \textit{TQC 2012}, pp.29-44.


\bibitem{nlwe99} C.H. Bennett, D.P. DiVincenzo, C.A. Fuchs, T. Mor, E. Rains, P.W. Shor, J.A. Smolin and W.K. Wootters, ``Quantum nonlocality without entanglement,'' \textit{Phys. Rev. A} 59, pp.1070–1091 (1999). 
\bibitem{nlwe12} A.M. Childs, D. Leung, L. Mancinska and M. Ozols, ``A framework for bounding nonlocality of state discrimination,'' arXiv:1206.5822.

\bibitem{qdh02} D.P. DiVincenzo, D.W. Leung and B.M. Terhal, ``Quantum Data Hiding,'' \textit{IEEE Trans. Inf. Theory}, Vol. 48, No. 3, pp.580-599 (2002).
\bibitem{eggeling02} T. Eggeling and R. F. Werner, ``Hiding Classical Data in Multipartite Quantum States,'' \textit{Phys. Rev. Lett.} 89, 097905 (2002).

\bibitem{masanes} L. Masanes, ``Universally Composable Privacy Amplification from Causality Constraints,'' \textit{Phys. Rev. Lett.} 102, 140501 (2009).
\bibitem{trevisan-vadhan-2000} L. Trevisan and S. P. Vadhan, ``Extracting Randomness from Samplable Distributions,'' \textit{FOCS} 2000, pp.32-42.
\bibitem{kamp-zuckerman-2006} J. Kamp and D. Zuckerman, ``Deterministic Extractors for Bit-Fixing Sources and Exposure-Resilient Cryptography,'' \textit{SIAM J. Comput.} Vol. 36, No. 5, pp. 1231–1247 (2006).
\bibitem{gabizon-2011} A. Gabizon, \textit{Deterministic Extraction from Weak Random Sources}, Springer-Verlag (2011).
\bibitem{AGV} A. Akavia, S. Goldwasser and V. Vaikuntanathan, ``Simultaneous Hardcore Bits and Cryptography against Memory Attacks,'' \textit{TCC 2009}, pp.474-495.
\bibitem{naor-segev} M. Naor and G. Segev, ``Public-Key Cryptosystems Resilient to Key Leakage,'' \textit{CRYPTO} 2009, pp.18-35.

\bibitem{damgard} I. Damgard, S. Fehr, R. Renner, L. Salvail and C. Schaffner, ``A Tight High-Order Entropic Quantum Uncertainty Relation with Applications,'' \textit{CRYPTO 2007}, pp.360-378.



\bibitem{horodecki} R. Horodecki, P. Horodecki, M. Horodecki and K. Horodecki, ``Quantum Entanglement,'' \textit{Rev. Mod. Phys.} 81, pp.865-942, 2009.

\end{thebibliography}
\end{document}